\documentclass[fleqn,usenatbib]{mnras}

\usepackage{newtxtext,newtxmath}

\usepackage[T1]{fontenc}
\usepackage{bm}

\DeclareRobustCommand{\VAN}[3]{#2}
\let\VANthebibliography\thebibliography
\def\thebibliography{\DeclareRobustCommand{\VAN}[3]{##3}\VANthebibliography}

\usepackage{graphicx}	
\usepackage{amsmath}	

\usepackage{hyperref}
\usepackage{multirow}
\usepackage{array}
\usepackage{orcidlink}
\usepackage{threeparttablex}

\title[Doppler Boosting of X-ray Binary Jets]{Observational Biases and Improved Modelling of Off-axis Relativistic Jets}

\author[A. J. Cooper et al.]{A. J. Cooper$^{1 \orcidlink{0000-0002-4033-3139}}$\thanks{E-mail: alexander.cooper@physics.ox.ac.uk},
A. P. Scott$^{1,2}$,
L. Rhodes$^{3,4 \orcidlink{0000-0003-2705-4941}}$,
F. Carotenuto$^{5 \orcidlink{0000-0002-0426-3276}}$,
A. K. Hughes$^{1 \orcidlink{0000-0003-0764-0687}}$,
J. H. Matthews$^{1 \orcidlink{0000-0002-3493-7737}}$, \newauthor
K. Savard$^{1 \orcidlink{0009-0001-8598-0639}}$,
F. J. Cowie$^{1 \orcidlink{0009-0009-0079-2419}}$, 
E. L. Elley$^{1 \orcidlink{0009-0002-5349-908X}}$,
C. Lilje$^{1 \orcidlink{0009-0004-1555-5356}}$,
R. Fender$^{1,6 \orcidlink{0000-0002-5654-2744}}$\\
$^{1}$Astrophysics, The University of Oxford, Keble Road, Oxford, OX1 3RH, UK\\
$^{2}$Astrophysics Research Centre, School of Mathematics and Physics, Queen's University Belfast, BT7 1NN, UK\\
$^{3}$Trottier Space Institute at McGill, 3550 Rue University, Montreal, Quebec H3A 2A7, Canada\\
$^{4}$Department of Physics, McGill University, 3600 Rue University, Montreal, Quebec H3A 2T8, Canada \\
$^{5}$INAF-Osservatorio Astronomico di Roma, Via Frascati 33, I-00078, Monte Porzio Catone (RM), Italy\\
$^{6}$Department of Astronomy, University of Cape Town, Private Bag X3, Rondebosch 7701, South Africa 
\\
}

\date{Accepted XXX. Received YYY; in original form ZZZ}

\pubyear{\the\year{}}

\begin{document}
\label{firstpage}
\pagerange{\pageref{firstpage}--\pageref{lastpage}}
\maketitle

\begin{abstract}
Relativistic Doppler boosting significantly affects the observed emission of astrophysical jets resulting in observational biases. In this work we investigate the observational biases and modelling opportunities which arise due to relativistic boosting using two X-ray binary case studies. Using the one-sided jet ejecta from MAXI J1535-571, we demonstrate that incorporating non-detections of the receding jet ejecta into kinematic modelling can significantly improve parameter estimation, reducing posterior uncertainties by over $40\%$. For the bipolar jets of MAXI J1820+070, we recover the intrinsic jet rest-frame emission of both approaching and receding jet components, demonstrating that they follow a common powerlaw evolution. Using this rest-frame emission profile as a base model, we show that current observational strategies strongly bias against detecting ejecta with high initial Lorentz factors $\gtrsim 5$ and receding ejecta components across a broad region of parameter space. These results highlight the importance of observational strategy selection, particularly early-time and late-time observations, and leveraging non-detections in the modelling of relativistic jets. More generally, quantifying observational biases and maximising modelling capabilities by incorporating the non-detection of receding jets can be employed to enhance interpretation of future gravitational-wave/optically-triggered observations of off-axis, extragalactic jetted transients. 
\end{abstract}

\begin{keywords}
ISM: jets and outflows -- X-rays: binaries -- radio continuum: transients
\end{keywords}



\section{Introduction}
Astrophysical jets are collimated outflows observed across orders of magnitude in energy and spatial scales. Typically, they are characterized as bipolar outflows powered by either continuous or transient accretion episodes leading to spatially connected or discrete ejecta respectively. The most powerful jets stem from accretion onto black holes (BHs): supermassive BHs at the centre of galaxies power the jets of active galactic nuclei (e.g. \citealt{blandford_2019}) and tidal disruption events (e.g., \citealt{gezari_2021}), newly born BHs likely power a subset of gamma-ray bursts (GRBs; e.g., \citealt{kumar_2015}), and accreting stellar-mass BHs power X-ray binary jets (XRBs; e.g., \citealt{fender+2006}). Astrophysical jets are thought to have a significant impact on the surrounding environment through feedback \citep{fabian_2012,gaspari_2013,soker16} and as sources of high-energy cosmic rays and neutrinos \citep{cooper_2020,stein_2021,van_velzen_2024,zhang_2025,bacon_2026}. Studying these jets enables an understanding of high-energy particle acceleration in shocks \citep{matthews_2020} and the nature of the jet-disc coupling of compact objects \citep{blandfordz_77,blandfordp_82}. 
\par
The observed electromagnetic emission of jets depends on the macrophysical jet properties, the microphysics of particle acceleration likely driven by internal and external shocks, and the Doppler boosting of flux with respect to the observer due to bulk relativistic motion. Doppler boosting depends only on the observer angle to the jet axis, $\theta_{\rm v}$, and the time-dependent bulk Lorentz factor, $\Gamma(t)$, of the emitting jet material. Boosting of emission is captured by the Doppler factor, $\delta$:
\begin{equation}
\label{eq:dopplerfactor}
    \delta = \big(\Gamma (1\mp\beta \cos\theta_{\rm v})\big)^{-1}
\end{equation}
where $\mp$ depends on whether the emitting region is for approaching ($-$) or receding ($+$). The observed flux from a jet is modified compared to emitted flux by: $F_{\nu, \rm obs} \approx F_{\nu, \rm em} \, \delta^{\kappa-\alpha}$, where $\alpha$ is the radiation spectral index such that $F_{\nu} \propto \nu^{\alpha}$ and $\kappa \sim 2-3$ (see subsection \ref{sect:initallorentzfactor1820} for further discussion). $\Gamma$ can differ vastly depending on the outflow considered \citep{matthews_2025}. Jets from young stellar objects can be sub-relativistic \citep{ray_2021}, whereas cosmological GRBs are initially ultra-relativistic \citep{ghirlanda_2018}. XRB jets are thought to be trans-relativistic \citep{Hjellming_1981,corbel_2002,wood_2021}, despite the fact that measuring the Lorentz factor of off-axis jets is notoriously difficult and subject to selection biases \citep{lilje_2026}. Novel instrumentation and transient search techniques mean that many more transient, off-axis extragalactic jets are likely to be discovered in the coming years. Firstly, off-axis GRBs originating from neutron star mergers will be detected through gravitational wave emission. The astrophysical jet associated with GW/GRB 170817 has an inclination angle of $\theta_{\rm v} \approx 15-20 \, \deg$ \citep{alexander_18,lamb_2020,mooley_2022_gw17} and there are hints that the receding jet may be contributing to late-time emission \citep{Dastidar_2024}. The next generation of gravitational wave observatories LIGO Voyager and 3G gravitational wave detectors (Cosmic Explorer and Einstein Telescope) are expected to detect $\sim10^{2}$ and $\sim10^{4}$ binary neutron star mergers per year, respectively \citep{2022A&A...667A...2S,2024PhRvD.110h3040B}. A subset of these sources will have associated off-axis jet emission, for which a detailed understanding of time-dependent Doppler boosting and how non-detections of the receding jet can be incorporated into modelling will be valuable. Moreover, blind transient searches across X-ray, optical, and radio frequencies are expected to detect off-axis or low $\Gamma_0$ GRBs, where the receding jet may significantly contribute to observed emission. Candidate off-axis GRBs have already been discovered at radio \citep{law_2018,mooley_2022,2025ApJ...995...61S,gulati_2026}, optical \citep{srinivasaragavan_2025,perley_2025}, and possibly X-ray \citep{2025NatAs...9..564L,2025ApJ...986..106G} frequencies. Future survey instruments such as LSST, DSA, and the SKA promise to detect many more orphan kilonovae/afterglows \citep{andreoni_19}, motivating a study of the biases and modelling opportunities presented by off-axis and receding jets. 
\par
This work focusses on the transient jet ejecta from Galactic BH-XRBs. XRBs are usually discovered when they go into X-ray outbursts initiated by accretion disc instabilities (e.g., \citealt{lasota_2001}), resulting in a jet viewing angle ($\theta_{\rm v}$) distribution which is consistent with being isotropic (e.g., the distribution of XRBs is uniform in $\cos\theta_{\rm v}$). In outburst, BH-XRBs are observed to produce large-scale transient ejecta associated with state changes in the accretion disc \citep{fender_2004}. The separation of large-scale\footnote{Large-scale is a subjective classification, but is loosely defined here as core-ejecta separations $\gtrsim 1$ arcsecond.} transient jet ejecta from the core source, which may be radio-loud due to a compact steady jet, can be resolved as a function of time via interferometric radio and high spatial resolution X-ray observations. Such behaviour has now been observed in 17 BH-XRBs \citep{mirabel_1994,Hjellming_1995, hjellming_2000,hannikainen_2001, corbel_2002, mioduszewski_2001, gallo_2004,corbel_2005,yang_2010,rushton_2017,russell_2019,miller-jones_2019, bright_2020, carotenuto_2021,williams_2022, wood_2023, bahramian_2023,zhang_2025}. 
However, both the approaching and receding component have been detected at radio frequencies for only 6 sources: GRS 1915+105 \citep{mirabel_1994},  GRO J1655--40 \citep{Hjellming_1995}, XTE J1550--564 \citep{hannikainen_2001}, H1743--322 \citep{corbel_2005}, MAXI J1820+070 \citep{bright_2020}, and MAXI J1848--015 \citep{bahramian_2023}. Sources with larger $\theta_{\rm v}$ and/or low bulk Lorentz factors result in Doppler factors close to unity. Therefore receding jets from these sources are less affected by deboosting of emission, and thus we may expect to detect a receding jet. It is widely assumed that both the receding and approaching jet, as well as the media into which they propagate, are similar, and thus non-detections of a receding jet component can be informative for jet modelling and parameter estimation. 
\par
Interestingly, there is indirect evidence that some XRB jet ejecta may reach up to ultra-relativistic $\Gamma \gtrsim 5$ Lorentz factors, known as Ultra-Relativistic Flows (URFs). \cite{fomalont_2001urf,fomalont_2001urf_b} conducted high spatial resolution observations of the neutron star (NS) XRB Scorpius X-1, in which bipolar radio lobes appeared to be energised by unseen $\Gamma > 3$ ejecta from the NS. \cite{motta_fender19_urf} reanalysed X-ray observations contemporaneous to the production of the hypothesised URFs (see also \citealt{2026MNRAS.546ag046S}), finding the appearance of quasi-periodic oscillations consistent with jet launching linking the unseen ejecta and the accretion flow. Finally, \cite{fender2004_urf} reported observations for another young NS-XRB Circinus X-1, where the unseen ejecta which energised lobes must have $\Gamma > 15$. Interestingly, this source is also the only NS-XRB to show significant core-ejecta separated jet ejecta \citep{cowie_2025}. The large inferred Lorentz factors of URFs imply significant deboosting of ejecta radiation. Given this, it is unclear if URFs are exclusively produced by neutron star XRBs. Such URFs could be ubiquitous or even common among XRBs, but remain undetected.
\par
In this work, we investigate how Doppler boosting affects observations and modelling of discrete ejecta from BH-XRBs. In Section \ref{sect:recedingjets}, we demonstrate how non-detections of receding jets can enhance kinematic modelling capabilities using a first case study source, the approaching jet component observed from MAXI J1535-571. In Section \ref{sect:maxi1820} we introduce our second case study source, the powerful double-sided jet ejecta from XRB MAXI J1820+070. We utilise kinematic fits to the data to recover the jet rest-frame flux of these ejecta, and show that they are consistent with having similar properties. We explore the $\Gamma_0 - \theta_{\rm v}$ parameter space, using the jets from MAXI J1820+070 as a baseline for intrinsic jet flux. We demonstrate that regions of this parameter space may result unobservable jets, and discuss observational strategies to maximise scientific return. We conclude with a brief discussion of the primary results and their implications in Section \ref{ref:discuss}.




\section{Utilizing Non-detections of Receding Jets: Case study of MAXI J1535}
\label{sect:recedingjets}
For the majority of  large-scale jet ejecta discoveries from XRBs, no receding jet ejecta is detected. In a number of cases, this is despite kinematic modelling implying a relatively large $\theta_{\rm v}$, where the receding ejecta may not be significantly deboosted. In this subsection, we investigate how kinematic modelling of the one-sided large-scale jet ejecta from MAXI J1535$-$571 (henceforth MAXI J1535) can be improved by taking into account the fact that the receding ejecta was not detected. In exploring this new technique, we make a number of simplifying assumptions aimed to demonstrate how this methodology can improve future kinematic modelling of off-axis jets.

\subsection{Physical constraints from non-detections}
MAXI J1535 is a BH-XRB which went into X-ray outburst in 2017 \citep{negoro_2017,nakahira_2018,Tao_2018}, and was observed across the electromagnetic spectrum from radio to X-rays \citep{huang_2018,russell_2017,miller_2018,parikh_2019,russell_2019,russell_2020,chauhan_2021}. The large-scale (approaching) jet ejection was tracked across the sky for nearly a year using the Australia Telescope Compact Array (ATCA) and MeerKAT. The jet launch date was estimated via kinematic modelling to be MJD $58017.4^{+4.0}_{-3.8}$ \citep{carotenuto_2024}, in agreement with X-ray observations of enhanced flux \citep{shang_2019} and the tentative detection of quasi-periodic oscillations \citep{stevens_2018}. HI absorption measurements derived a best-fit distance to the source of $4.1^{+0.6}_{-0.5}$ kpc, with an upper limit of $6.7$ kpc and a lower limit of $3.5\,$kpc \citep{chauhan_2019}. A comprehensive discussion of the source and jet ejecta is presented in \cite{cooper_2025}, in which the authors presented combined Bayesian lightcurve and kinematic modelling of the jet ejecta. This approach found that both a reverse and forward shock are required to explain the observed lightcurves, and derived best-fit parameters for the preferred jet model (Model A in their work) of: $\Gamma_0 = 1.39 \pm 0.02$, $E_{0, \rm min} = 3.5^{+9.7}_{-2.4} \times 10^{43} \; {\rm erg}$,  $n_0 = 4.0^{+11.2}_{-2.7} \times 10^{-5} \; {\rm cm^{-3}}$, and $\theta_{\rm v} = 71.70 \pm 2.37 \deg$. Crucially for this work, the derived parameters imply a relatively small Doppler factor throughout the jet evolution, due to a combination of a low $\Gamma_0$ and relatively high $\theta_{\rm v}$.\footnote{We note that these results depend primarily on the flux modelling, supplemented by kinematic modelling, and thus are unaffected by degeneracies in $\Gamma_0$ discussed in the following paragraphs.} This makes the source a good choice to investigate whether the non-detection of a receding jet can aid in modelling, as the receding jet should be relatively bright due to a small Doppler factor.
\par
First, we perform a new kinematic-only fit to the approaching jet separation data presented in \cite{russell_2019}. For this, we opt to use the \texttt{jetsimpy}\footnote{\href{https://jetsimpy.readthedocs.io/en/latest/}{https://jetsimpy.readthedocs.io}} \citep{wang_2024} package to ensure consistency with previous studies of this source \citep{cooper_2025}. \texttt{jetsimpy} is an efficient reduced hydrodynamic code to model the kinematics and radiation of astrophysical jets, offering a balance of speed, flexibility, and fidelity. Similarly to \cite{cooper_2025}, we employ the nested sampling \citep{Skilling_2004} package \texttt{dynesty}\footnote{\href{https://dynesty.readthedocs.io/en/stable/}{https://dynesty.readthedocs.io/en/stable/}} \citep{speagle_2020}. We employ a standard Gaussian likelihood function:
\begin{equation}
    \ln \mathcal{L}(\theta) = -\frac{n}{2} {\rm log}(\sigma^2) - \frac{1}{2} \sum_{i=1}^{n} \bigg(\frac{x_i - \mu}{\sigma}\bigg)^2
    \label{eq:reg_loglike}
\end{equation}
where $x_{i}$ are the model separations, $\mu$ and $\sigma$ are the core-ejecta separation data and errors respectively. We use $1024$ live points, random walk sampling (\texttt{rwalk}), and the default multi-ellipsoidal decomposition (\texttt{multi}) and a stopping criterion of \texttt{dlogz = $0.001$}, where \texttt{dlogz} is the log of the ratio between the current estimated evidence and the remaining evidence.
\par
\begin{figure}
  \centering
{\includegraphics[width=.5\textwidth]{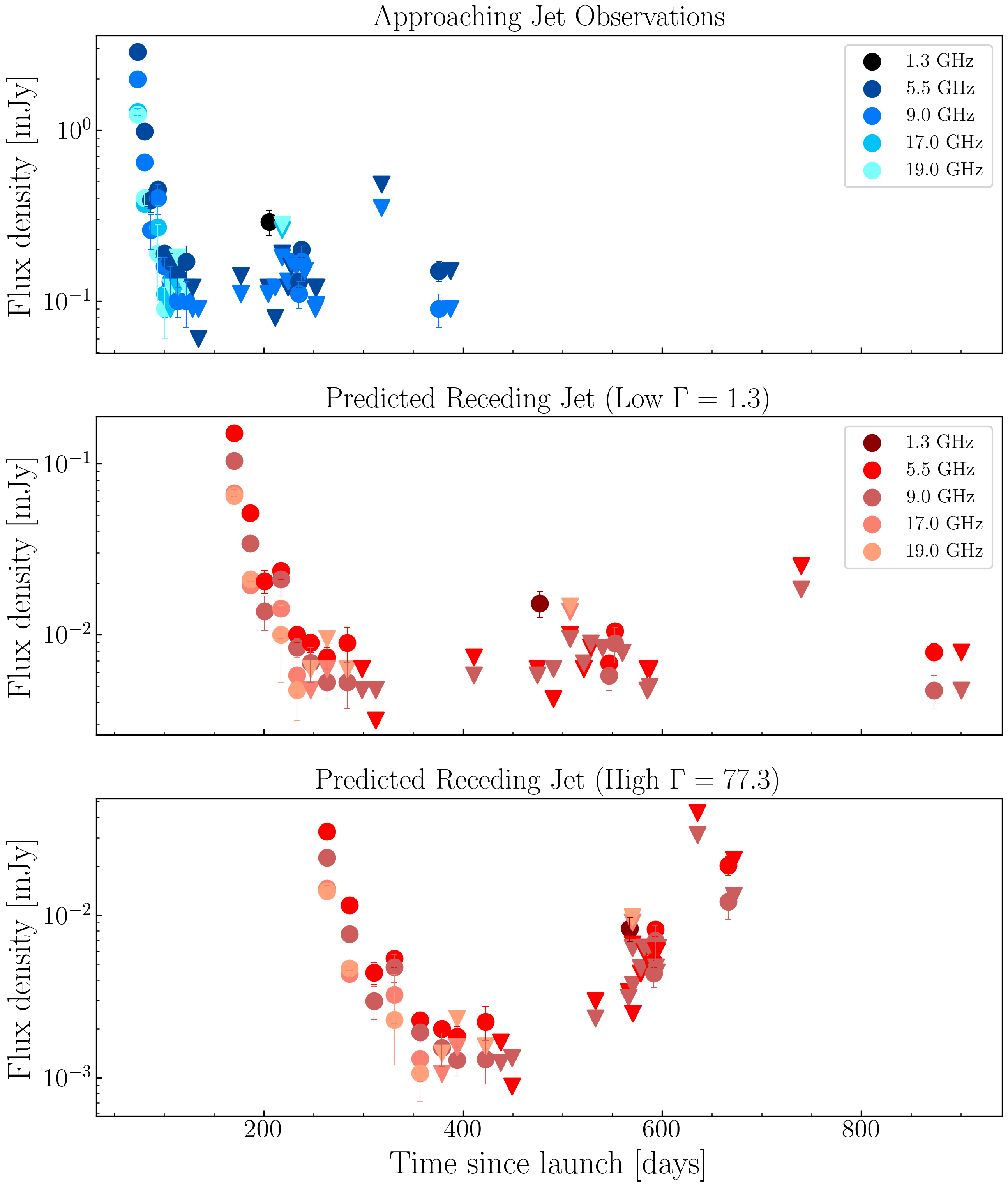}}
\caption{Approaching (top panel) and predicted receding (bottom two panels) jet flux from the \textit{initial} kinematic fit of MAXI J1535, coloured by frequency. As expected, the receding jet flux is lower and delayed as compared to approaching jet flux for both low-$\Gamma_0$ and high-$\Gamma_0$ solutions. The underlying data are published in \protect\cite{russell_2019}.}
\label{fig:predictedflux}
\end{figure}

To circumvent known degeneracies associated with kinematic-only fitting (see e.g., \citealt{carotenuto_2024}), we initially fix some parameters to those obtained in \cite{cooper_2025}: the half-opening angle $\theta_{\rm c} = 2.25 \deg$ and the circumsource density of $n_0 = 4.3 \times 10^{-5} \, {\rm cm^{-3}}$, where we note that low circumsource density appears to be a hallmark of XRBs, possibly carved out by previous jet activity \citep{carotenuto_2024,savard_2025}. The initial energy $E_0$ (where $E_{0} = E_{\rm iso} \theta_{\rm c}^2/16$), Lorentz factor $\Gamma_0$, observer viewing angle $\theta_{\rm v}$, and the distance $d_{\rm kpc}$ are the free parameters. The priors on these free parameters are given in Table \ref{tab:priors}. Fig. \ref{fig:predictedflux} shows example predicted receding jet lightcurves and in Fig. \ref{fig:corner1} we show the posterior corner plot of this first kinematic fit. The use of nested sampling in place of MCMC in \cite{carotenuto_2024}, who performed kinematic fits to this source, results in very different posterior distributions for $\Gamma_0$. We find that $\Gamma_0$ is essentially unconstrained for kinematic-only modelling (see also \citealt{lilje_2026}), and determining this parameter requires additional information to be incorporated in the fit, such as the ejecta flux \citep{cooper_2025}. A full study of $\Gamma_0$ fitting with kinematic models, including differences between nested sampling and MCMC, will be presented in a future work.


\setlength{\tabcolsep}{8pt}
\setlength{\extrarowheight}{.7em}
\begin{table*}
\caption{List of parameters and priors for the kinematic fits of MAXI J1535}
\label{tab:priors}
\begin{tabular}{*{4}{c l c c}}
\hline
\hline
\textbf{Parameter}  & \textbf{Description} 	&   \textbf{Prior} & \textbf{Bounds/Value}\\
\hline                                  
$\Gamma_0$ & Initial bulk Lorentz factor & Uniform & [1,100]\\
${E}_{\rm iso}$ & Isotropic-Equivalent Initial Energy [erg] & Log-uniform & [$10^{40}$,\,$10^{50}$] \\
$\theta_{\rm v}$ & Jet Viewing Angle [degrees] & Cosine & [$0$,$90$] \\
$D$     & Source distance [kpc] &  Normal [Truncated] & $4.1_{-0.6}^{+0.6}$ [$1$,\,$8$]\\
$n_0$ & Circumsource Density [${\rm cm}^{-3}$]& \textbf{Fixed} & \bm{$10^{-4.37}$}\\
$\theta_{\rm c}$ & Half-opening angle of jet [degrees] & \textbf{Fixed} & \textbf{2.25} \\
$t_{\rm ej}$ & Launch time of ejecta & \textbf{Fixed} & \textbf{MJD 58017.4}\\
\hline         
\end{tabular}
\end{table*}

\par
We fit the source again with the same priors and sampling parameters, but now incorporating the fact that no receding jet ejecta was detected. For each sampling step we compute the deceleration and Doppler factor profile of the approaching jet $\Gamma(t_{\rm obs})$ and $\delta(t_{\rm obs})$, which allows us to convert observed approaching jet flux datapoints into receding jet predictions. First, we convert approaching jet fluxes presented in \cite{russell_2019} to the rest frame using the following equation:
\begin{equation}
    F_{\nu, \rm rest} = F_{\nu, \rm obs}/\delta^{3-\alpha}
    \label{eq:flux_corrected}
\end{equation}
where $\alpha$ is the spectral index. We apply the same equation to go from the rest-frame to the receding jet frame, using the Doppler profile of the receding jet for the given parameters. Note that the exponent of $3-\alpha$ utilised in Eq. \ref{eq:flux_corrected} is thought to be appropriate for discrete ejecta \citep{1985ApJ...295..358L}, however the real value may vary between $2-3$ (see e.g., \citealt{1995PASP..107..803U}), and has been estimated as $\sim2.3$ for jet ejecta observed from GRS 1915+105 (\citealt{mirabel_1994}; see also \citealt{fender_1999} who estimate a lower value of 1.3-1.9 for subsequent later ejecta). Next, we correct the time-of-arrival of the predicted receding jet fluxes. The time-varying jet velocity and Doppler factor $\delta(t_{\rm obs})$ of decelerating ejecta requires an integration to move from the observers view of the approaching jet to the jet proper time (e.g. jet comoving or rest frame, $\tau_{\rm jet}$):
\begin{equation}
    \tau_{\rm jet} = \int_0^{t_{\rm obs}^{\rm app}} \frac{dt'}{\delta_{\rm app}(t')}
        \label{eq:temporal_correction}
\end{equation}
We can then invert this operation to move from the jet rest frame to the observer's view of the receding jet:
\begin{equation}
    t_{\rm obs}^{\rm rec} = \int_0^{\tau_{\rm jet}} \delta_{\rm rec}(\tau')\, d\tau'
\end{equation}


In Fig. \ref{fig:predictedflux}, we show the observed data for the approaching jet (top panel) and the predicted receding jet flux for two draws from the posterior parameters at low and high $\Gamma_0$ values respectively (bottom two panels), each of which fit the kinematic data equally well. Within our new framework, the highest flux data point in the second panel low-$\Gamma_0$ solution would result in a penalty to the likelihood. 

\par
To incorporate the information gained due to non-detections of the receding jet, we penalise the likelihood if the receding could have been observed for each datapoint. This allows us to better constrain the jet parameters, as some parameters should have resulted in receding jets bright enough to detect. For this test case, we do not require that real-world observations of MAXI J1535 were conducted at the time of the synthesised receding jet datapoint. We simply implement two `observability criteria' required to be met by the synthesised receding jet flux datapoints. First, the predicted flux density must be above an observable threshold which we set in terms of the signal-to-noise of the ejecta flux as $\rm{SNR} = F_{\rm ejecta}/F_{\rm noise}  > 8$, adopting a flat noise value of $F_{\rm noise} = 10 \, \mu {\rm Jy}$, typical of a 30 minute MeerKAT observation. Second, it must be significantly separated from the core such that it would be spatially distinguished from the core.
Following Eq. 51 in \cite{Lobanov_2005}, we define the limiting separation as:
\begin{equation}
    \theta_{\rm lim} = 2^{2-\frac{\beta}{2}} b \Bigg[ \frac{\ln 2}{\pi} \ln\bigg(\frac{{\rm SNR}}{\rm{SNR -1}}\bigg) \Bigg]^{1/2}
    \label{eq:snr_rs}
\end{equation}
Here, $\theta_{\rm lim}$ is the threshold resolution along the axis separating the ejecta and the core, $b$ is the size of the beam (full width half-maximum) along the same axis, and $\beta$ is the weighting for which we assume $\beta=2$ corresponding to natural weighting. For simplicity, we assume that the beam size is constant, adopting $b = 5$ arcseconds typical of 1.28 GHz MeerKAT or 5 GHz ATCA observations. If these criteria are met, we implement a penalty to the likelihood based on the fact that the receding jet should have been seen, for a given parameter set:
\begin{equation}
\mathcal{L}_{\rm pen}(\theta)
= \ln \mathcal{L}(\theta)
- \sum_{i\in\mathcal{I}}
\left[
\max\!\left(0,\frac{f_i^{\rm rec}(\theta)-f_i^{\rm thr}}{f_i^{\rm thr}}\right)
\right]^2
\end{equation}
where $f_i^{\rm rec}(\theta)$ is the predicted flux for a given set of parameters, $f_i^{\rm thr}$ is the observability threshold, and $\mathcal{L}(\theta)$ is the likelihood defined in Eq. \ref{eq:reg_loglike}.

\subsection{Results for MAXI J1535}

\begin{figure}
  \centering
{\includegraphics[width=.5\textwidth]{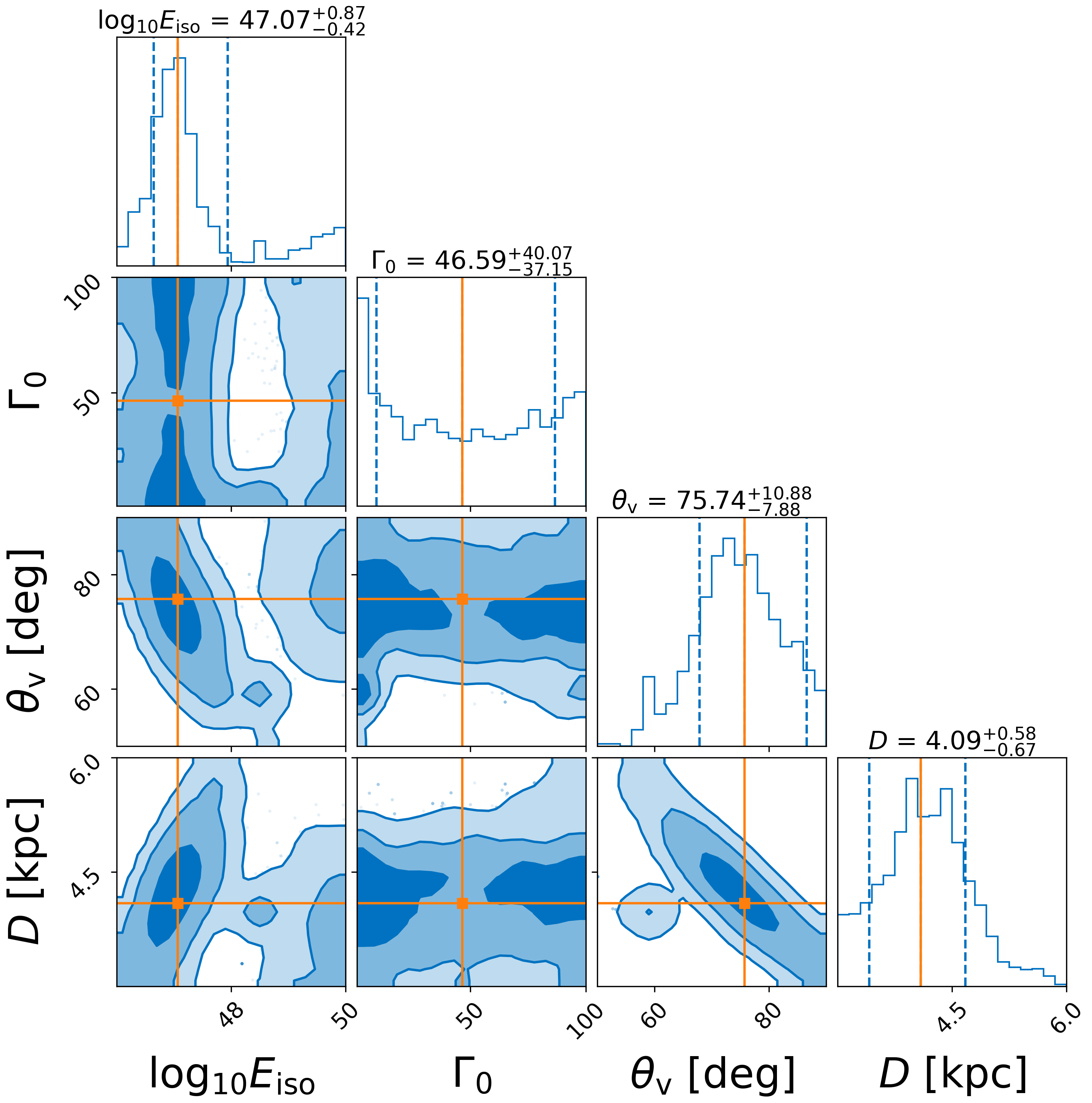}}
\caption{Corner plot of posterior parameter space of kinematic-only fit to the separation data of MAXI J1535 \textit{without} receding jet constraints. Orange lines show the best-fit parameters determined by the maximum likelihood, and blue contours highlight 1$\sigma$, 2$\sigma$, and 3$\sigma$ credible regions. Here, $E_{\rm iso} = 16 E_{0}/\theta_{\rm c}^2 \approx 10^4 E_0$ for our fixed value of $\theta_{\rm c} = 2.25 \deg$.}
\label{fig:corner1}
\end{figure}
\begin{figure}
  \centering
{\includegraphics[width=.5\textwidth]{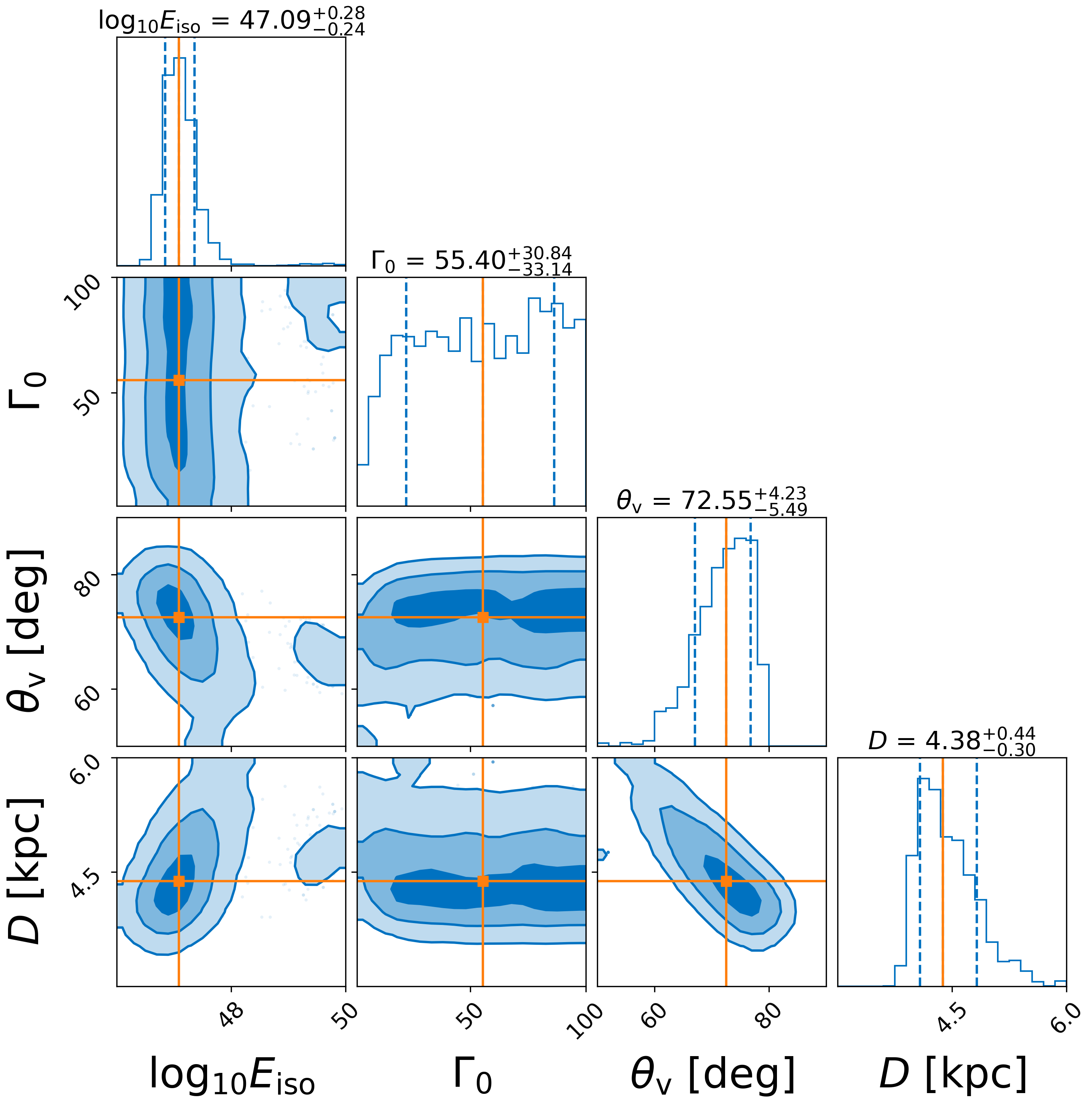}}
\caption{Corner plot of posterior parameter space of kinematic-only fit to MAXI J1535 \textit{including} receding jet constraints.}
\label{fig:corner2}
\end{figure}

We find that the estimated parameters are roughly consistent across both the initial kinematic fit (Fig. \ref{fig:corner1}) and the second one which incorporates constraints from the non-detection of the receding jet (Fig. \ref{fig:corner2}). The initial Lorentz factor is relatively unconstrained in both cases, but a significant low-$\Gamma$ cut-off is observed in the second, receding jet constrained fit. This low-$\Gamma$ cut-off is expected as $\Gamma \to 1$ would imply a receding component with identical flux evolution to the approaching component, which should have been detected. A sharp high $\theta_{\rm v}$ cut-off in the posterior is observed for similar reasons. This in turn enforces a definitive low distance cut-off in the posterior of the second fit, as the separation $\alpha_{\rm sep} \propto \sin(\theta_{\rm v})/D$. Inspection of the posterior parameter space shows that the high-energy peak in the bimodal posterior distribution of $\log(E_{\rm iso)}$ occurs as some kinematic solutions completely ignore the final separation datapoint at $\sim 400$ days post-launch. This bimodal behaviour is suppressed significantly when receding constraints are taken into account. Overall, we find that incorporating receding ejecta constraints for this source significantly improves parameter estimation, reducing the 1$\sigma$ uncertainty by an average of \textbf{41.5\%} across all four parameters, increasing to \textbf{49.7\%} when the initial Lorentz factor is excluded. We stress that precise reductions in parameter estimation uncertainty are specific to this source and depend strongly on the viewing angle to the source, as this methodology is best-suited to jet ejecta with large $\theta_{\rm v}$. 
 
\section{Exploring the $\Gamma_0$-$\theta_{\rm v}$ parameter space: Case study of MAXI J1820}
\label{sect:maxi1820}
MAXI J1820+070 (henceforth MAXI J1820) is a BH-XRB discovered in the optical \citep{tucker_2018} in March 2018 by the All-Sky Automated Survey for SuperNovae (ASAS-SN), and a week later in X-ray \citep{Kawamuro_2018} by the Monitor of All-sky X-ray Image (MAXI). Neutron Star Interior Composition Explorer (NICER) observations identified MAXI J1820 as a candidate BH via timing properties \citep{uttley_2018}, before dynamical BH confirmation \citep{Torres_2019}. \cite{atri_2020} obtained a precise radio parallax distance to the source of $2.96\pm0.33$ kpc utilising Very Long Baseline Interferometry (VLBI). \cite{bright_2020} presented radio observations of jet ejecta stemming from the 2018 outburst of MAXI J1820, where the core-ejecta separation was resolved for both the approaching and receding jet components. X-ray emission from the large-scale discrete ejecta was observed and presented in \citep{espinasse_2020}, and further early-time core-separation datapoints were recovered by \cite{wood_2021}, as well as additional slower ejecta, utilising dynamic phase-centre-tracked VLBI observations. 
\par
\begin{figure}
  \centering
{\includegraphics[width=.5\textwidth]{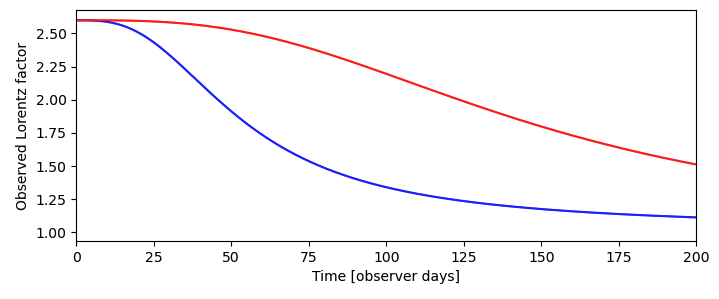}}
\caption{Lorentz factor observer-frame temporal profile of the approaching (blue) and receding (red) components of MAXI J1820 jet ejecta \protect\citep{bright_2020} from kinematic modelling presented in \protect\cite{carotenuto_2024}.}
\label{fig:decel_curves_1820}
\end{figure}

The kinematic data of both jets was fit by \cite{carotenuto_2024} utilising a trans-relativistic, conical blastwave model (e.g. constant half-opening angle, $\phi_{\rm jet}$), in which the ejecta shell decelerates as material is swept up. This kinematic model, which is similar in nature to the approach in the previous section utilising \texttt{jetsimpy}, is discussed in full in previous works \citep{huang_1999,wang_2003,steiner_2012,carotenuto_2022a,carotenuto_2024}. In brief, the model requires solving for the angular separation of the ejecta, for a source at a distance $D$:
\begin{equation}
    \alpha_{\rm sep}(t) = \frac{R(t) \sin(\theta_{\rm v})}{D}
\end{equation}
by integrating the kinematic equation \citep{rees_1966,mirabel_1994}:
\begin{equation}
    \frac{dR}{dt} = \frac{\beta c}{1 \mp \beta \cos\theta_{\rm v}}
\end{equation}
where $\beta = v_{\rm jet}/c$ is the normalised jet velocity, and $\mp$ corresponds to approaching ($-$) and receding ($+$) jets respectively. To obtain a solution for $\Gamma(t)$, we assume the total ejecta energy is conserved \citep{huang_1999,wang_2003} such that:
\begin{equation}
    E_0 = (\Gamma(t) - 1) M_0 c^{2} + \sigma (\Gamma_{\rm sh}(t)^2 - 1) m_{\rm sw}(t) c^2
\end{equation}
Here $\Gamma_{\rm sh}$ is the Lorentz factor of the shock front, $m_{\rm sw} \propto n_0 \phi^2 R^3$ is the mass swept by the shock, and $\sigma$ is a numerical factor to interpolate between non-relativistic and relativistic shocks \citep{huang_1999}. Kinematic modelling is enhanced for MAXI J1820, as the approaching and receding components can be jointly fit assuming identical jets and deceleration profiles. \cite{carotenuto_2024} derive an initial Lorentz factor of $\Gamma_0 = 2.6^{+0.5}_{-0.4}$, a viewing angle of $\theta_{\rm v} = 59.6^{+1.0}_{-1.2}$ and an `effective energy' (e.g., the $E_0$ obtained when fixing $n_0 = 1 \, {\rm cm^{-3}}$ and $\phi_{\rm jet} = 1\deg$) of $E_{\rm eff} = 2.6^{+ 0.4}_{-0.4} \times 10^{46} \, {\rm erg}$. Only an `effective energy' can be derived due the aforementioned degeneracy between $E_{\rm 0}$, $n_0$, and $\phi_{\rm jet}$ in kinematic-only modelling \citep{carotenuto_2021}. In Fig. \ref{fig:decel_curves_1820} we show the best fit deceleration $\Gamma(t_{\rm obs})$ profiles from \cite{carotenuto_2024}. These kinematic profiles are used in subsection \ref{sect:recovering} to recover the jet frame emission. 

\subsection{The initial Lorentz factor of the large-scale ejecta from MAXI J1820}
\label{sect:initallorentzfactor1820}
The posterior distribution of the initial Lorentz factor obtained through kinematic modelling is sensitive to the choice of sampler in the fitting. To test this, we perform a new fit to the double-sided jets of MAXI J1820, utilising the same data as \cite{carotenuto_2024}. All parameter choices and priors are kept the same, but the the nested sampling package \texttt{dynesty} is used in place of \texttt{emcee}\footnote{https://emcee.readthedocs.io/en/stable/} \citep{foreman_mackey_2013}. In Appendix Fig. \ref{fig:1820_corner}, we show the new parameter posteriors, which are in good agreement with those derived by \cite{carotenuto_2024} with the exception of $\Gamma_0$ which is poorly constrained. However, the detection of double-sided jet ejecta provides additional information which aids us in discriminating between the large range of $\Gamma_0$ values. Approaching and receding jets at the same spatial separation are of the same intrinsic age, despite reaching these separations at different times in the observer frame \citep{fender_2003}. One can use this fact to compare same-separation flux densities ($F_{\nu}$) using the following equation:
\begin{equation}
    F_{\nu, \rm rec} \approx F_{\nu, \rm app} \delta_{\rm rec}^{3-\alpha}/\delta_{\rm app}^{3-\alpha}
    \label{eq:fluxComparison}
\end{equation}
where $\alpha = -0.5$ is the spectral index and $\delta$ is the Doppler factor of each jet at the same separation (Eq. \ref{eq:dopplerfactor}). Taking a fixed value of the angle to the line of sight of $\theta_{\rm v} \sim 59 \deg$ \citep{carotenuto_2024}, we can solve for $\Gamma(r_{\rm sep})$ a distance $r_{\rm sep}$ by comparing $F_{\nu, \rm rec}(r_{\rm sep})$ and $F_{\nu, \rm app}(r_{\rm sep})$. 
\par
No strictly `same-separation' radio detections exist for this source, however the closest radio datapoints have measured separations of $6.4''\pm0.8$ and $5.4''\pm0.4$ arcseconds for the approaching and receding jet respectively. These occur on MJD $58389$ and MJD $58484$, with measured fluxes of $2.26\pm0.05\,$mJy and $0.10\pm0.008\,$mJy at 1.28 GHz and 6.0 GHz for the approaching and receding jet respectively. Extrapolation to exact same intrinsic age is not possible, as the approaching jet has no clear radio detections before this time except at early-times prior to the main flux increase, and the receding jet has no radio detections at later times. Nonetheless, we can still compare these datapoints to gain some insight into $\Gamma(r_{\rm sep})$. Assuming a spectral index $\alpha = - 0.5$ \citep{bright_2020}, the approaching jet flux is approximately $1.04\,$mJy at 6.0 GHz. This rough estimate suggests via Eq. \ref{eq:fluxComparison}, a Lorentz factor of $\Gamma(r_{\rm sep}) \approx 1.284^{+0.37}_{-0.31}$. The quoted uncertainty here arises from the propagating reported flux errors alone and does not account for the difference in separation (and intrinsic age) of the jets. Nonetheless despite this additional uncertainty, the derived value matches well with the deceleration profiles of the best-fit model of \cite{carotenuto_2024} shown in Fig. \ref{fig:decel_curves_1820}, providing confidence in their best-fit value of $\Gamma_0 \approx 2.6$. In the following, we proceed with analysis of the MAXI J1820 dataset based on the best-fit parameters of \cite{carotenuto_2024}, but note that further work is required to fully constrain $\Gamma_0$ for this source.

\subsection{Recovering the rest-frame emission of MAXI J1820}
\label{sect:recovering}
Utilising the best fit kinematic profile, we have all the information required to obtain the rest-frame flux evolution of each jet, by effectively `deboosting' the observed lightcurves, similarly to Section \ref{sect:recedingjets}. First, the flux is scaled to a common frequency of $1.28$ GHz, assuming $F_{\nu} \propto \nu^{\alpha}$ where we take $\alpha = -0.5$. Next, the flux density is corrected to account for Doppler boosting using Eq. \ref{eq:flux_corrected}. Finally, we correct the observer time to the rest frame time by inverting Eq. \ref{eq:temporal_correction}.
\begin{figure}
  \centering
{\includegraphics[width=.5\textwidth]{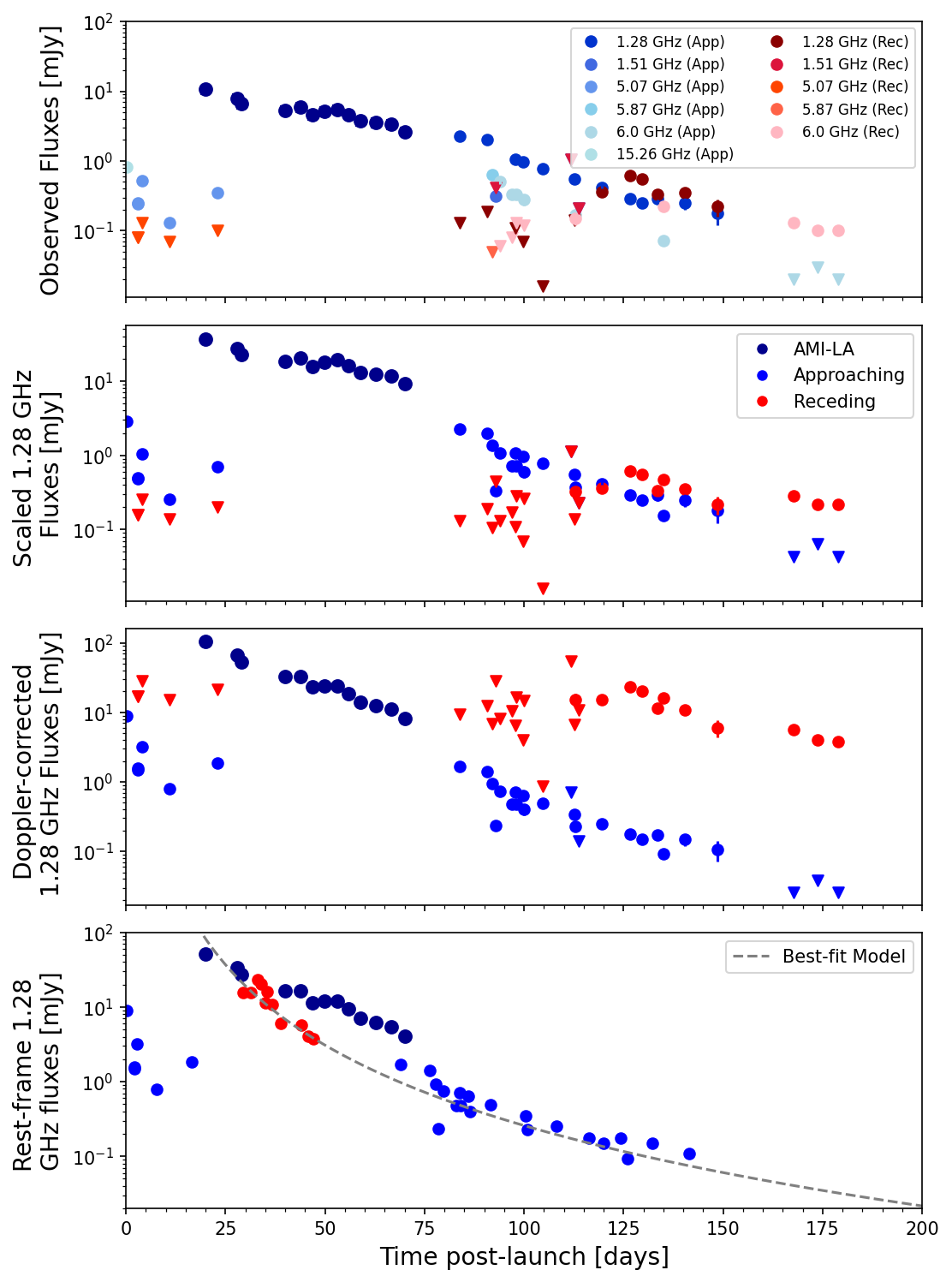}}
\caption{Panel 1: Observed fluxes of approaching jet (blue colours), receding jet (red colours), and unresolved AMI-LA (dark blue) from MAXI J1820. Panel 2: Same data scaled to a common frequency of 1.28 GHz. Panel 3: Same data corrected for (de)boosting of the flux due to bulk relativistic motion assuming the decelerating blast wave model fit. Panel 4: Same detections corrected to rest-frame emission time, with the simple powerlaw model in grey (see text).}
\label{fig:fluxes_1820}
\end{figure}

In Fig. \ref{fig:fluxes_1820} we show the radio fluxes of both jet components presented in \cite{bright_2020} and \cite{wood_2021}. Panel 1 shows the raw data in the observer frame, coloured by frequency, where blue shades indicate flux from the approaching jet and red shades indicate observations of the receding jet. The early-time $t<25$ days datapoints are VLBI data with e-Merlin (5 GHz) and VLBA (15.26 GHz) which are associated with the approaching ejecta \citep{bright_2020,wood_2021}. The darkest blue AMI-LA datapoints are unresolved from the core and thus their origin (core, approaching, or receding jet, or some combination of these three components) cannot be robustly determined. Nevertheless, we naively corrected the AMI-LA data to the rest frame by assuming it is dominated by the approaching jet, but do not use these data for the fit. In the second panel, all detections and upper limits are scaled to a common frequency of 1.28 GHz. In the third panel, the fluxes and upper limits are corrected for Doppler boosting using Eq. \ref{eq:flux_corrected}. Finally in the fourth panel, all detections  are corrected to the rest frame using Eq. \ref{eq:temporal_correction}.\footnote{In comparing rest-frame deceleration curves, we find a small disagreement (always $<5\%$) between the approaching and receding jets, likely due to a numerical error in the integration. This does not significantly affect our results.}
\par
A number of results are immediately apparent. Firstly, both the approaching and receding jet detections appear to obey a common powerlaw in their respective rest frames, with the exception of the earliest datapoints. While we assumed jet symmetry to transform observed flux evolution to the rest-frame, this still supports long-held assumptions of intrinsically similar jet evolution as similar evolution does not follow from kinematic fit a priori. The fact that the earliest datapoints at $t < 25$ days do not follow the powerlaw flux decay can be interpreted as evidence that the reverse shock, which is thought to power particle acceleration in XRB ejecta \citep{wang_2003,savard_2025,cooper_2025,matthews_2025}, has not yet crossed the ejecta. We caution that these detections are at high spatial resolution and may resolve out a portion of the flux. Secondly, despite being detected weeks later by the observer, the receding jet component emission was likely emitted much earlier and at significantly higher intrinsic luminosities than the observed approaching jet component. While this is expected, this clearly demonstrates how detecting the receding component of relativistic outflows can provide crucial information to gain a full picture of the temporal evolution of the blastwave.
\par
To study the XRB jet ejecta population more broadly, we fit a simple powerlaw to the rest-frame emission of the approaching and receding components of MAXI J1820, excluding the early-time high spatial resolution and unresolved AMI-LA datapoints. We employ the functional form:
\begin{equation}
    F_{\nu, \rm rest} = S_{0} \bigg(\frac{t}{t_0}\bigg)^{\gamma}
\end{equation}
We opt to use $t_0 = 25$ days, the jet rest frame time corresponding to the earliest clear detection of the receding jet. We utilise the \texttt{emcee} package to sample the parameter space, using conservative, uniform priors of $-3 < S_0 < 10$, $-10 < \gamma < 1$, $-10 < \log(f) < 1$. $\log(f)$ is a fudge factor we fit for to ensure a few datapoints with unreasonably small errors bars do not distort the fit. We find best-fit values of $\log_{10}(S_0) = 1.566\pm0.07$ and $\gamma = -3.579^{+0.13}_{-0.14}$, which is plotted in the bottom panel of Fig. \ref{fig:fluxes_1820} as a grey dashed line. 

\subsection{Are the fastest XRB jets unobservable?}
\label{sect:fastestjets}
To study the wider $\Gamma_0-\theta_{\rm v}$ XRB jet ejecta parameter space, we utilise the best-fit rest frame powerlaw flux evolution as a base model. We make predictions for similar emission for arbitrary initial Lorentz factors ($\Gamma_0$) and observer angles ($\theta_{\rm v})$. We suppress the model at early times by enforcing zero flux at $t_{\rm rest} < 25 $ days, as the early-time emission of MAXI J1820 does not follow the powerlaw. As aforementioned, this is physically motivated as the reverse shock is thought to cross the ejecta only tens of days post-launch \citep{matthews_2025,cooper_2025}, depending on the initial blastwave energy, $E_0$, the density of the surrounding medium, $n_0$, and $\Gamma_0$. Enforcing an earlier time cut-off results in larger predicted fluxes, particularly for the receding jet.

\par
First, we compute kinematic profiles, $\Gamma(t_{\rm obs})$ and $\alpha_{\rm sep}(t_{\rm obs})$ (on-sky separation) for jets with $1.1 \leq \Gamma_0 \leq 11.1$ and $0 < \theta_{\rm v} < \pi/2$, where $\theta_{\rm v}$ is expressed in radians. We keep the effective energy $E_{\rm eff}$ as the best fit value from \cite{carotenuto_2024}. We then transform the rest-frame flux model using these kinematic profiles to obtain synthetic MAXI J1820-like lightcurves for arbitrary $\Gamma_0$ and $\theta_{\rm v}$. This is done by applying Doppler boosting and photon time-of-arrival corrections using Eqs. \ref{eq:flux_corrected} \& \ref{eq:temporal_correction} to calculate $F_{\nu, \rm obs}$ from $F_{\nu, \rm rest}$ for both approaching and receding jets.
\par
\begin{figure}
  \centering
{\includegraphics[width=.5\textwidth]{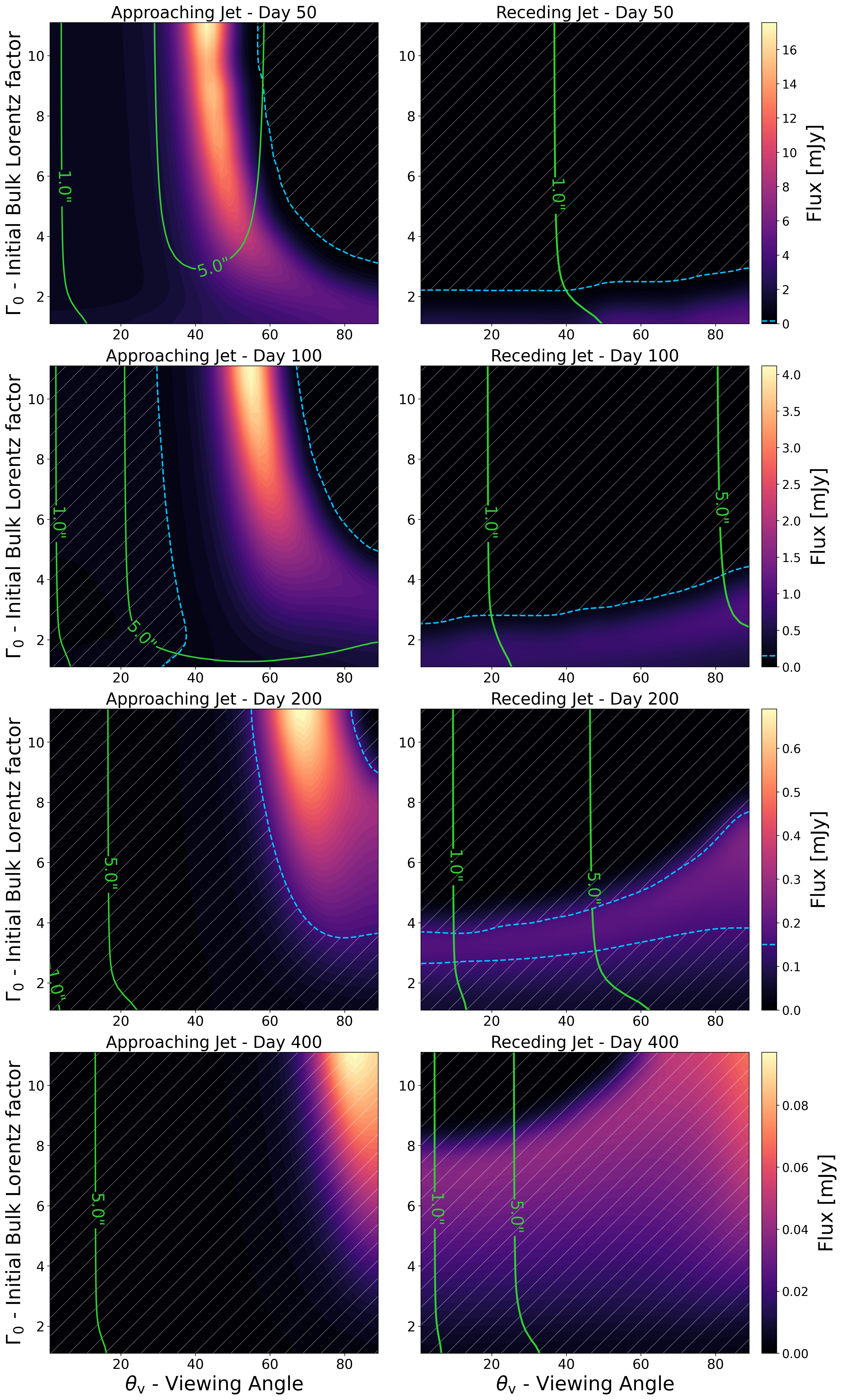}}
\caption{Predicted MAXI J1820-like emission from the approaching (left panels) and receding (right panels) as a function of the initial Lorentz factor $\Gamma_0$ and viewing angle $\theta_{\rm v}$ for 50, 100, 200, and 400 days post-launch in the observer frame. Note that the colour scale changes for each row. Cyan dashed contours correspond to $F_{\nu} = 150 \, {\rm \mu Jy}$ flux, with the white hatched region excluding regions of lower flux. The green contours correspond to core-ejecta separations of 1 and 5 arcseconds. Generally, approaching jets with small $\theta_{\rm v}$ can only be detected at early times, with the minimum viewing angle detectable increasing over time. Receding jets are generally limited to low-$\Gamma$ ejecta due to deboosting, although the band of observable emission moves to higher $\Gamma$ at later times. At late times, all jet emission is too dim to be detected.}
\label{fig:flux_day_100_500}
\end{figure}

In Fig. \ref{fig:flux_day_100_500}, we show the expected 1.28 GHz flux for both receding and approaching 1820-like jets across the $\Gamma_0-\theta_{\rm v}$ parameter space at 50, 100, 200, and 400 days post-launch. Note that the changing colourbar reflects the fact that both jets' fluxes are decreasing with time, as expected by the intrinsic powerlaw flux model. As a sanity check, the rightmost column of each plot ($\theta_{\rm v} = 90 \deg$) is common to both approaching and receding jets, corresponding to jets in the plane of the sky where the jets are identical for the observer, albeit with slight discontinuities due to limited numerical resolution. The dashed cyan contours indicate fluxes bounded by $F_{\nu} > 150 \, {\rm \mu Jy}$, such that all regions not shaded with white hatches are detectable at this level. The green contours demarcate a jet-core separation of $1''$ and $5''$ arcseconds, corresponding to a typical range of angular resolutions of GHz radio interferometers.
\par
Interpretation of the flux evolution of the jet is non-trivial, as the profile depends on both the intrinsic powerlaw flux decay, and the time-varying Doppler factor which can be positive or negative. The approaching jet displays a band of maximal flux which decays in flux and moves towards higher inclination angle jets at later times. Early-time observations are essential to detect the approaching jet of low inclination angle systems (see also Fig. \ref{fig:observability}), but high inclination angle systems ($\theta_{\rm v} > 60 \deg$; which is the majority of systems) require late-time observations after 100 days post-launch. This is particularly important when $\Gamma_0\gtrsim4$ and is plausibly the reason for the non-detection of aforementioned URFs, despite the differences in compact objects between XRBs with URFs and those discussed in this work. 
\par
For the receding jet, emission is heavily suppressed for higher Lorentz factors due to deboosting, as expected. For modest Lorentz factors of $2 < \Gamma_0 < 5$, large parts of the parameter space require late-time observations $100 \:{\rm days} \lesssim t \lesssim 300 \:{\rm days}$ for the detection of the receding component, often months after the approaching ejecta has faded below the sensitivity threshold. Moreover, high-spatial resolution and sensitive observations ($\lesssim 10 \mu $Jy/beam noise levels) are crucial to detect the receding component and resolve it from the core. Most troublesome is that, in the low-$\Gamma_0$, low-$\theta_{\rm v}$ parameter space, observations after 50 days may only detect a receding component, which could be confused with either the approaching jet component or the jet core. Generically, for higher initial Lorentz factors ($\Gamma_0 > 2$), only large viewing angles $\theta_{\rm v} \gtrsim 45$ permit the approaching and receding jet to be simultaneously detectable. 





\begin{figure}
  \centering
{\includegraphics[width=.5\textwidth]{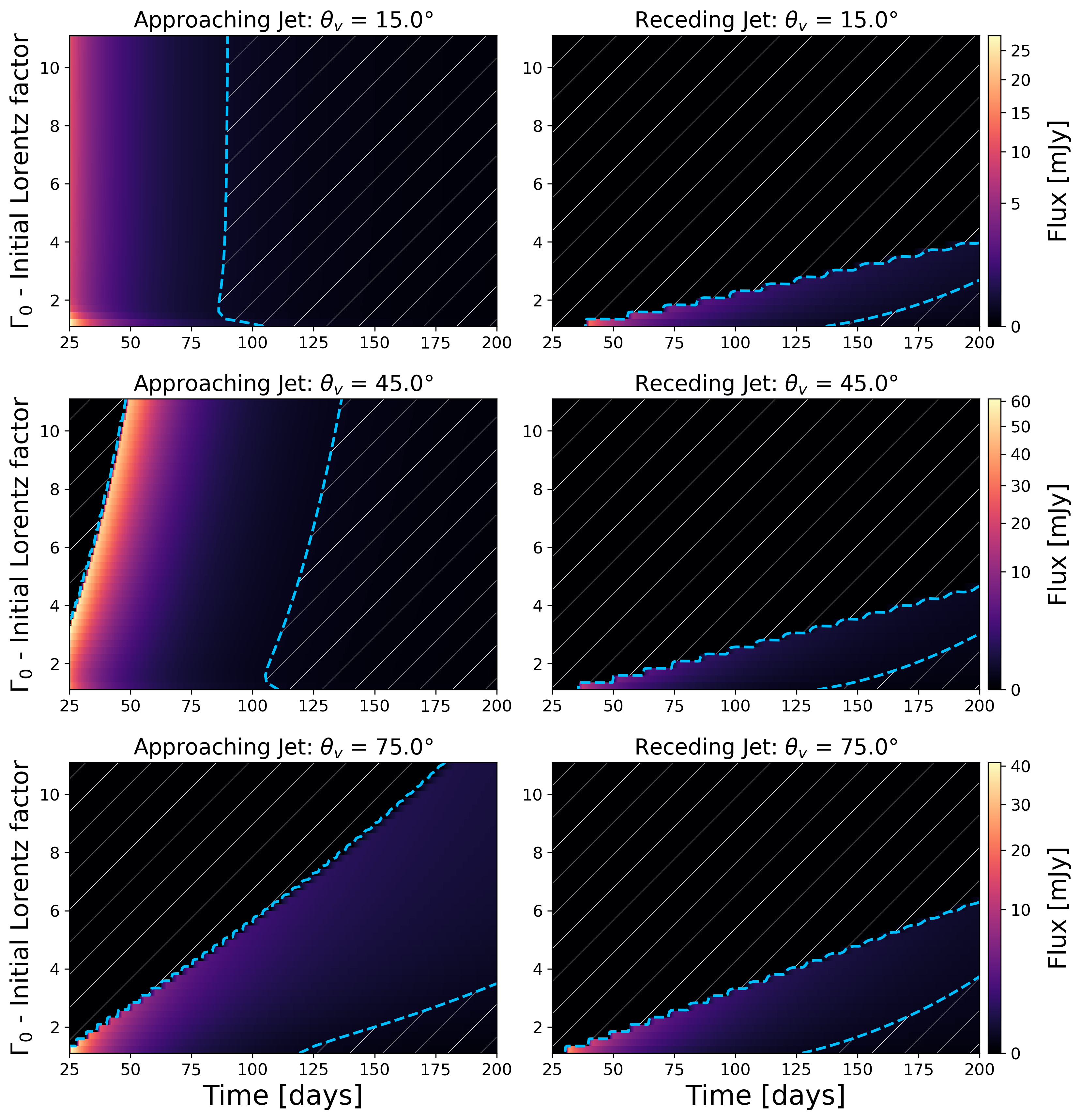}}
\caption{Predicted MAXI J1820-like emission from the approaching (left panels) and receding (right panels) as a function of time and $\Gamma_0$ for three viewing angles ($\theta_{\rm v}$ = [15, 45, 75] $\deg$). Note the powerlaw colour scale changes for each row. Cyan dashed contours again correspond to $F_{\nu} = 150 \, {\rm \mu Jy}$ flux, with the white hatched region excluding regions of lower flux. The shape of the observable region shows how observations at different times are sensitive to ejecta with differing initial Lorentz factors in a manner that strongly depends on the viewing angle. }
\label{fig:flux_over_time_difftheta}
\end{figure}

In Fig. \ref{fig:flux_over_time_difftheta} we show the flux evolution as a function of different initial Lorentz factors for three viewing angles. The time-axis runs from day 25 to day 200 jet post-launch in the observer frame, demonstrating the difficulty associated with detecting receding jets with high $\Gamma_0$. Prior constraints on $\theta_{\rm v}$, combined with a detection at a particular time, can indicate a range of $\Gamma_0$. However, this is much more easily achieved when the receding ejecta is detected. In the Appendix, we show in Fig. \ref{fig:observability} the parameter space for jet detectability on day 60, 120, 200, and 280 days post-launch, adopting a 1.28 GHz detection threshold of $0.15$ mJy, roughly $\gtrsim 5 \sigma$ given the typical RMS noise of the 15-minute MeerKAT scans common in BH-XRB datasets.

\par





\section{Conclusions}
\label{ref:discuss}
In this work, we have investigated the Doppler boosting of large-scale XRB jet ejecta, and the implications this has for optimising observing strategies to reduce biases in our understanding of XRB jets. We have demonstrated, using a case study of MAXI J1535, that in cases where prior information implies a large $\theta_{\rm v}$, non-detections of the receding jet ejecta can significantly improve parameter estimation through kinematic modelling. By recovering the intrinsic jet rest frame emission of MAXI J1820 as a case study, we have predicted emission from similar jet ejecta across different initial Lorentz factors and observer angles. We have demonstrated that difficulties in detecting ejecta across the parameter space, particularly for high $\Gamma_0$ jets, may introduce significant observational bias in the inferred distribution of XRB jet ejecta Lorentz factors (see also \citealt{lilje_2026}). Such bias can be partially overcome by prioritising high spatial resolution, early-time observations at $10$s days post-launch, and sensitive late-time observations at $>200$ days post-launch. The primary conclusions of this work can be summarised as follows. 
\begin{enumerate}
    \item Incorporating the non-detection of the receding jet in kinematic fits of large-scale ejecta can significantly improve parameter estimation, reducing uncertainty by $40\%$ for MAXI J1535. \\
    \item The large-scale jet ejecta from MAXI J1820 are consistent with an intrinsic powerlaw flux of $F_{\nu} \propto t^{-3.6}$, with receding ejecta flux being produced much earlier in the jet rest frame despite arriving to the observer at much later times.\\
    \item Approaching high-$\Gamma_0$ jet ejecta can go undetected unless both early-time and late-time observations are conducted. \\
    \item Sensitive, high-spatial resolution observations are required at late times ($\gtrsim$ 200 days) to detect the receding ejecta of even moderate-$\Gamma_0$ ejecta. For much of the parameter space, this component is only visible weeks or months after the approaching jet has faded below detectable levels. 
\end{enumerate}

To maximise scientific output from multi-wavelength, kinematically resolved observations of X-ray binary ejecta, all available data should be taken into account during modelling. In particular, two separate approaches should be considered. Firstly, kinematic-only fitting, which offers a less model-dependent approach, should fold in constraints due to the non-detections of the receding jet or flux mismatches at similar intrinsic ages for receding jet detections. The second approach involves jointly fitting the kinematics and additional data such as the jet ejecta lightcurves \citep{cooper_2025}. Future works may also incorporate polarimetric and ejecta size constraints through likelihood penalisations as presented here, and compare models to realistic simulations \citep{savard_2025}. 
\par
Finally, while our case studies have focussed on XRB jet ejecta, both the physics and statistical methodology presented in this work are widely applicable to jetted transients. In particular, the modelling of future off-axis relativistic, extragalactic jets, including gravitational-wave triggered or optically triggered GRBs and tidal disruption events, may benefit from employing similar approaches.

\section*{Acknowledgements}

AJC acknowledges support from the Oxford Hintze Centre for Astrophysical Surveys which is funded through generous support from the Hintze Family Charitable Foundation and the Oxford UNIQ+ Summer School. LR acknowledges funding from the  Trottier Space Institute Fellowship, the Natural Sciences and Engineering Research Council of Canada (NSERC) Arthur B. McDonald Fellowship and Discovery Grant programs, the Canada Research Chairs (CRC) program, the Fondes de Recherche Nature et Technologies (FRQNT), the Centre de recherche en astrophysique du Québec (un regroupement stratégique du FRQNT), and the AstroFlash research group. The AstroFlash research group at McGill University, University of Amsterdam, ASTRON, and JIVE is supported by: a Canada Excellence Research Chair in Transient Astrophysics (CERC-2022-00009); an Advanced Grant from the European Research Council (ERC) under the European Union's Horizon 2020 research and innovation programme (`EuroFlash'; Grant agreement No. 101098079); an NWO-Vici grant (`AstroFlash'; VI.C.192.045); an NSERC Discovery Grant (RGPIN-2025-06681); an ERC Starting Grant (`EnviroFlash'; Grant agreement No. 101223057); and an NWO-Veni grant (VI.Veni.222.295). JHM and EE acknowledges funding from a Royal Society University Research Fellowship (URF\textbackslash R1\textbackslash221062). FJC is supported by STFC grant ST/Y509474/1. RF acknowledges support from UK Research and Innovation, The European Research Council, and the Hintze Family Charitable Foundation. We gratefully acknowledge the use of the following software packages: \texttt{matplotlib} \citep{matplotlib}, \texttt{emcee} \citep{foreman_mackey_2013}, and \texttt{dynesty} \citep{speagle_2020}. 


\section*{Data Availability}
All observational data referenced in this work is freely available in the referenced publications, but can be provided upon request to the corresponding author. 

\bibliographystyle{mnras}
\bibliography{example}

@ARTICLE{wang_2024,
       author = {{Wang}, Hao and {Dastidar}, Ranadeep G. and {Giannios}, Dimitrios and {Duffell}, Paul C.},
        title = "{jetsimpy: A Highly Efficient Hydrodynamic Code for Gamma-Ray Burst Afterglow}",
      journal = {\apjs},
         year = 2024,
        month = jul,
       volume = {273},
       number = {1},
          eid = {17},
        pages = {17},
          doi = {10.3847/1538-4365/ad4d9d},
archivePrefix = {arXiv},
       eprint = {2402.19359},
 primaryClass = {astro-ph.HE},
       adsurl = {https://ui.adsabs.harvard.edu/abs/2024ApJS..273...17W}
}

@ARTICLE{fender_2003,
       author = {{Fender}, R.~P.},
        title = "{Uses and limitations of relativistic jet proper motions: lessons from Galactic microquasars}",
      journal = {\mnras},
         year = 2003,
        month = apr,
       volume = {340},
       number = {4},
        pages = {1353-1358},
          doi = {10.1046/j.1365-8711.2003.06386.x},
archivePrefix = {arXiv},
       eprint = {astro-ph/0301225},
 primaryClass = {astro-ph},
       adsurl = {https://ui.adsabs.harvard.edu/abs/2003MNRAS.340.1353F}
}

@ARTICLE{matplotlib,
       author = {{Hunter}, John D.},
        title = "{Matplotlib: A 2D Graphics Environment}",
      journal = {Computing in Science and Engineering},
         year = 2007,
        month = jan,
       volume = {9},
       number = {3},
        pages = {90-95},
          doi = {10.1109/MCSE.2007.55},
       adsurl = {https://ui.adsabs.harvard.edu/abs/2007CSE.....9...90H}
}

@ARTICLE{1985ApJ...295..358L,
       author = {{Lind}, K.~R. and {Blandford}, R.~D.},
        title = "{Semidynamical models of radio jets: relativistic beaming and source counts.}",
      journal = {\apj},
         year = 1985,
        month = aug,
       volume = {295},
        pages = {358-367},
          doi = {10.1086/163380},
       adsurl = {https://ui.adsabs.harvard.edu/abs/1985ApJ...295..358L}
}

@ARTICLE{2022A&A...667A...2S,
       author = {{Singh}, Neha and {Bulik}, Tomasz and {Belczynski}, Krzysztof and {Askar}, Abbas},
        title = "{Exploring compact binary populations with the Einstein Telescope}",
      journal = {\aap},
         year = 2022,
        month = nov,
       volume = {667},
          eid = {A2},
        pages = {A2},
          doi = {10.1051/0004-6361/202142856},
archivePrefix = {arXiv},
       eprint = {2112.04058},
 primaryClass = {astro-ph.HE},
       adsurl = {https://ui.adsabs.harvard.edu/abs/2022A&A...667A...2S}
}

@ARTICLE{2024PhRvD.110h3040B,
       author = {{Borhanian}, Ssohrab and {Sathyaprakash}, B.~S.},
        title = "{Listening to the Universe with next generation ground-based gravitational-wave detectors}",
      journal = {\prd},
         year = 2024,
        month = oct,
       volume = {110},
       number = {8},
          eid = {083040},
        pages = {083040},
          doi = {10.1103/PhysRevD.110.083040},
archivePrefix = {arXiv},
       eprint = {2202.11048},
 primaryClass = {gr-qc},
       adsurl = {https://ui.adsabs.harvard.edu/abs/2024PhRvD.110h3040B}
}

@ARTICLE{Dastidar_2024,
       author = {{Dastidar}, Ranadeep G. and {Duffell}, Paul C.},
        title = "{Could the Recent Rebrightening of the GW170817A Afterglow Be Caused by a Counterjet?}",
      journal = {\apj},
         year = 2024,
        month = dec,
       volume = {976},
       number = {2},
          eid = {252},
        pages = {252},
          doi = {10.3847/1538-4357/ad86bf},
archivePrefix = {arXiv},
       eprint = {2410.12185},
 primaryClass = {astro-ph.HE},
       adsurl = {https://ui.adsabs.harvard.edu/abs/2024ApJ...976..252D}
}

@ARTICLE{alexander_18,
       author = {{Alexander}, K.~D. and {Margutti}, R. and {Blanchard}, P.~K. and {Fong}, W. and {Berger}, E. and {Hajela}, A. and {Eftekhari}, T. and {Chornock}, R. and {Cowperthwaite}, P.~S. and {Giannios}, D. and {Guidorzi}, C. and {Kathirgamaraju}, A. and {MacFadyen}, A. and {Metzger}, B.~D. and {Nicholl}, M. and {Sironi}, L. and {Villar}, V.~A. and {Williams}, P.~K.~G. and {Xie}, X. and {Zrake}, J.},
        title = "{A Decline in the X-Ray through Radio Emission from GW170817 Continues to Support an Off-axis Structured Jet}",
      journal = {\apjl},
         year = 2018,
        month = aug,
       volume = {863},
       number = {2},
          eid = {L18},
        pages = {L18},
          doi = {10.3847/2041-8213/aad637},
archivePrefix = {arXiv},
       eprint = {1805.02870},
 primaryClass = {astro-ph.HE},
       adsurl = {https://ui.adsabs.harvard.edu/abs/2018ApJ...863L..18A}
}

@ARTICLE{lamb_2020,
       author = {{Lamb}, Gavin P. and {Levan}, Andrew J. and {Tanvir}, Nial R.},
        title = "{GRB 170817A as a Refreshed Shock Afterglow Viewed Off-axis}",
      journal = {\apj},
         year = 2020,
        month = aug,
       volume = {899},
       number = {2},
          eid = {105},
        pages = {105},
          doi = {10.3847/1538-4357/aba75a},
archivePrefix = {arXiv},
       eprint = {2005.12426},
 primaryClass = {astro-ph.HE},
       adsurl = {https://ui.adsabs.harvard.edu/abs/2020ApJ...899..105L}
}

@ARTICLE{mooley_2022_gw17,
       author = {{Mooley}, Kunal P. and {Anderson}, Jay and {Lu}, Wenbin},
        title = "{Optical superluminal motion measurement in the neutron-star merger GW170817}",
      journal = {\nat},
         year = 2022,
        month = oct,
       volume = {610},
       number = {7931},
        pages = {273-276},
          doi = {10.1038/s41586-022-05145-7},
archivePrefix = {arXiv},
       eprint = {2210.06568},
 primaryClass = {astro-ph.HE},
       adsurl = {https://ui.adsabs.harvard.edu/abs/2022Natur.610..273M}
}

@ARTICLE{mooley_2022,
       author = {{Mooley}, K.~P. and {Margalit}, B. and {Law}, C.~J. and {Perley}, D.~A. and {Deller}, A.~T. and {Lazio}, T.~J.~W. and {Bietenholz}, M.~F. and {Shimwell}, T. and {Intema}, H.~T. and {Gaensler}, B.~M. and {Metzger}, B.~D. and {Dong}, D.~Z. and {Hallinan}, G. and {Ofek}, E.~O. and {Sironi}, L.},
        title = "{Late-time Evolution and Modeling of the Off-axis Gamma-Ray Burst Candidate FIRST J141918.9+394036}",
      journal = {\apj},
         year = 2022,
        month = jan,
       volume = {924},
       number = {1},
          eid = {16},
        pages = {16},
          doi = {10.3847/1538-4357/ac3330},
archivePrefix = {arXiv},
       eprint = {2107.04703},
 primaryClass = {astro-ph.HE},
       adsurl = {https://ui.adsabs.harvard.edu/abs/2022ApJ...924...16M}
}

@ARTICLE{law_2018,
       author = {{Law}, C.~J. and {Gaensler}, B.~M. and {Metzger}, B.~D. and {Ofek}, E.~O. and {Sironi}, L.},
        title = "{Discovery of the Luminous, Decades-long, Extragalactic Radio Transient FIRST J141918.9+394036}",
      journal = {\apjl},
         year = 2018,
        month = oct,
       volume = {866},
       number = {2},
          eid = {L22},
        pages = {L22},
          doi = {10.3847/2041-8213/aae5f3},
archivePrefix = {arXiv},
       eprint = {1808.08964},
 primaryClass = {astro-ph.HE},
       adsurl = {https://ui.adsabs.harvard.edu/abs/2018ApJ...866L..22L}
}

@ARTICLE{perley_2025,
       author = {{Perley}, Daniel A. and {Ho}, Anna Y.~Q. and {Fausnaugh}, Michael and {Lamb}, Gavin P. and {Kasliwal}, Mansi M. and {Ahumada}, Tomas and {Anand}, Shreya and {Andreoni}, Igor and {Bellm}, Eric and {Bhalerao}, Varun and {Bolin}, Bryce and {Brink}, Thomas G. and {Burns}, Eric and {Cenko}, S. Bradley and {Corsi}, Alessandra and {Filippenko}, Alexei V. and {Frederiks}, Dmitry and {Goldstein}, Adam and {Hamburg}, Rachel and {Jayaraman}, Rahul and {Jonker}, Peter G. and {Kool}, Erik C. and {Kulkarni}, Shrinivas R. and {Kumar}, Harsh and {Laher}, Russ and {Levan}, Andrew and {Lysenko}, Alexandra and {Perley}, Richard A. and {Ricker}, George R. and {Riddle}, Reed and {Ridnaia}, Anna and {Rusholme}, Ben and {Smith}, Roger and {Svinkin}, Dmitry and {Ulanov}, Mikhail and {Vanderspek}, Roland and {Waratkar}, Gaurav and {Yao}, Yuhan},
        title = "{The luminous, slow-rising orphan afterglow AT2019pim as a candidate moderately relativistic outflow}",
      journal = {\mnras},
         year = 2025,
        month = mar,
       volume = {537},
       number = {3},
        pages = {2362-2379},
          doi = {10.1093/mnras/staf125},
archivePrefix = {arXiv},
       eprint = {2401.16470},
 primaryClass = {astro-ph.HE},
       adsurl = {https://ui.adsabs.harvard.edu/abs/2025MNRAS.537.2362P}
}

@ARTICLE{srinivasaragavan_2025,
       author = {{Srinivasaragavan}, Gokul P. and {Perley}, Daniel A. and {Ho}, Anna Y.~Q. and {O'Connor}, Brendan and {de Ugarte Postigo}, Antonio and {Sarin}, Nikhil and {Cenko}, S. Bradley and {Sollerman}, Jesper and {Rhodes}, Lauren and {Green}, David A. and {Svinkin}, Dmitry S. and {Bhalerao}, Varun and {Waratkar}, Gaurav and {Nayana}, A.~J. and {Chandra}, Poonam and {Miller}, M. Coleman and {Malesani}, Daniele B. and {Ryan}, Geoffrey and {Srijan}, Suryansh and {Bellm}, Eric C. and {Burns}, Eric and {Titterington}, David J. and {Stone}, Maria B. and {Purdum}, Josiah and {Ahumada}, Tom{\'a}s and {Anupama}, G.~C. and {Barway}, Sudhanshu and {Coughlin}, Michael W. and {Drake}, Andrew and {Fender}, Rob and {Ag{\"u}{\'\i} Fern{\'a}ndez}, Jos{\'e} F. and {Frederiks}, Dmitry D. and {Geier}, Stefan and {Graham}, Matthew J. and {Kasliwal}, Mansi M. and {Kulkarni}, S.~R. and {Kumar}, Harsh and {Li}, Maggie L. and {Laher}, Russ R. and {Lysenko}, Alexandra L. and {Parwani}, Gopal and {Perley}, Richard A. and {Ridnaia}, Anna V. and {Salgundi}, Anirudh and {Smith}, Roger and {Sravan}, Niharika and {Swain}, Vishwajeet and {Th{\"o}ne}, Christina C. and {Tsvetkova}, Anastasia E. and {Ulanov}, Mikhail V. and {Vail}, Jada and {Wise}, Jacob L. and {Wold}, Avery},
        title = "{Multiwavelength analysis of AT 2023sva: a luminous orphan afterglow with evidence for a structured jet}",
      journal = {\mnras},
         year = 2025,
        month = mar,
       volume = {538},
       number = {1},
        pages = {351-372},
          doi = {10.1093/mnras/staf290},
archivePrefix = {arXiv},
       eprint = {2501.03337},
 primaryClass = {astro-ph.HE},
       adsurl = {https://ui.adsabs.harvard.edu/abs/2025MNRAS.538..351S}
}

@ARTICLE{andreoni_19,
       author = {{Andreoni}, Igor and {Anand}, Shreya and {Bianco}, Federica B. and {Cenko}, S. Bradley and {Cowperthwaite}, Philip S. and {Coughlin}, Michael W. and {Drout}, Maria and {Golkhou}, V. Zach and {Kaplan}, David L. and {Mooley}, Kunal P. and {Pritchard}, Tyler A. and {Singer}, Leo P. and {Webb}, Sara and {LSST Transient}, with support of the and {Variable Stars Collaboration}},
        title = "{A Strategy for LSST to Unveil a Population of Kilonovae without Gravitational-wave Triggers}",
      journal = {\pasp},
         year = 2019,
        month = jun,
       volume = {131},
       number = {1000},
        pages = {068004},
          doi = {10.1088/1538-3873/ab1531},
archivePrefix = {arXiv},
       eprint = {1812.03161},
 primaryClass = {astro-ph.IM},
       adsurl = {https://ui.adsabs.harvard.edu/abs/2019PASP..131f8004A}
}

@ARTICLE{gulati_2026,
       author = {{Gulati}, Ashna and {Murphy}, Tara and {Kaplan}, David L. and {Dobie}, Dougal and {Ward}, Charlotte and {Anderson}, Gemma and {Caleb}, Manisha and {Chandra}, Poonam and {Cooke}, Jeff and {Das}, Barnali and {Deller}, Adam and {Goodwin}, Adelle and {Gourdji}, Kelly and {Ghirlanda}, Giancarlo and {Lenc}, Emil and {M{\"o}ller}, Anais and {Leung}, James K. and {Ocker}, Stella Koch and {Pritchard}, Joshua and {Ricci}, Claudio and {Sadler}, Elaine M. and {Sharan Salafia}, Om and {Shaji}, Kavya and {Soria}, Roberto and {Suhr}, Mark and {Tuntsov}, Artem and {Wang}, Ziteng},
        title = "{ASKAP J005512.2-255834: A Luminous, Long-Lived Radio Transient at z = 0.1 -- an Orphan Afterglow or an off-nuclear TDE from an IMBH?}",
      journal = {arXiv e-prints},
         year = 2026,
        month = feb,
          eid = {arXiv:2602.20522},
        pages = {arXiv:2602.20522},
          doi = {10.48550/arXiv.2602.20522},
archivePrefix = {arXiv},
       eprint = {2602.20522},
 primaryClass = {astro-ph.HE},
       adsurl = {https://ui.adsabs.harvard.edu/abs/2026arXiv260220522G}
}

@ARTICLE{1995PASP..107..803U,
       author = {{Urry}, C. Megan and {Padovani}, Paolo},
        title = "{Unified Schemes for Radio-Loud Active Galactic Nuclei}",
      journal = {\pasp},
         year = 1995,
        month = sep,
       volume = {107},
        pages = {803},
          doi = {10.1086/133630},
archivePrefix = {arXiv},
       eprint = {astro-ph/9506063},
 primaryClass = {astro-ph},
       adsurl = {https://ui.adsabs.harvard.edu/abs/1995PASP..107..803U}
}

@ARTICLE{tucker_2018,
       author = {{Tucker}, M.~A. and {Shappee}, B.~J. and {Holoien}, T.~W.-S. and {Auchettl}, K. and {Strader}, J. and {Stanek}, K.~Z. and {Kochanek}, C.~S. and {Bahramian}, A. and {ASAS-SN} and {Dong}, Subo and {Prieto}, J.~L. and {Shields}, J. and {Thompson}, Todd A. and {Beacom}, John F. and {Chomiuk}, L. and {ATLAS} and {Denneau}, L. and {Flewelling}, H. and {Heinze}, A.~N. and {Smith}, K.~W. and {Stalder}, B. and {Tonry}, J.~L. and {Weiland}, H. and {Rest}, A. and {Huber}, M.~E. and {Rowan}, D.~M. and {Dage}, K.},
        title = "{ASASSN-18ey: The Rise of a New Black Hole X-Ray Binary}",
      journal = {\apjl},
         year = 2018,
        month = nov,
       volume = {867},
       number = {1},
          eid = {L9},
        pages = {L9},
          doi = {10.3847/2041-8213/aae88a},
archivePrefix = {arXiv},
       eprint = {1808.07875},
 primaryClass = {astro-ph.HE},
       adsurl = {https://ui.adsabs.harvard.edu/abs/2018ApJ...867L...9T}
}

@ARTICLE{gaspari_2013,
       author = {{Gaspari}, M. and {Ruszkowski}, M. and {Oh}, S. Peng},
        title = "{Chaotic cold accretion on to black holes}",
      journal = {\mnras},
         year = 2013,
        month = jul,
       volume = {432},
       number = {4},
        pages = {3401-3422},
          doi = {10.1093/mnras/stt692},
archivePrefix = {arXiv},
       eprint = {1301.3130},
 primaryClass = {astro-ph.CO},
       adsurl = {https://ui.adsabs.harvard.edu/abs/2013MNRAS.432.3401G}
}

@ARTICLE{fabian_2012,
       author = {{Fabian}, A.~C.},
        title = "{Observational Evidence of Active Galactic Nuclei Feedback}",
      journal = {\araa},
         year = 2012,
        month = sep,
       volume = {50},
        pages = {455-489},
          doi = {10.1146/annurev-astro-081811-125521},
archivePrefix = {arXiv},
       eprint = {1204.4114},
 primaryClass = {astro-ph.CO},
       adsurl = {https://ui.adsabs.harvard.edu/abs/2012ARA&A..50..455F}
}

@ARTICLE{van_velzen_2024,
       author = {{van Velzen}, Sjoert and {Stein}, Robert and {Gilfanov}, Marat and {Kowalski}, Marek and {Hayasaki}, Kimitake and {Reusch}, Simeon and {Yao}, Yuhan and {Garrappa}, Simone and {Franckowiak}, Anna and {Gezari}, Suvi and {Nordin}, Jakob and {Fremling}, Christoffer and {Sharma}, Yashvi and {Yan}, Lin and {Kool}, Erik C. and {Stern}, Daniel and {Veres}, Patrik M. and {Sollerman}, Jesper and {Medvedev}, Pavel and {Sunyaev}, Rashid and {Bellm}, Eric C. and {Dekany}, Richard G. and {Duev}, Dimitri A. and {Graham}, Matthew J. and {Kasliwal}, Mansi M. and {Kulkarni}, Shrinivas R. and {Laher}, Russ R. and {Riddle}, Reed L. and {Rusholme}, Ben},
        title = "{Establishing accretion flares from supermassive black holes as a source of high-energy neutrinos}",
      journal = {\mnras},
         year = 2024,
        month = apr,
       volume = {529},
       number = {3},
        pages = {2559-2576},
          doi = {10.1093/mnras/stae610},
archivePrefix = {arXiv},
       eprint = {2111.09391},
 primaryClass = {astro-ph.HE},
       adsurl = {https://ui.adsabs.harvard.edu/abs/2024MNRAS.529.2559V}
}

@ARTICLE{stein_2021,
       author = {{Stein}, Robert and {van Velzen}, Sjoert and {Kowalski}, Marek and {Franckowiak}, Anna and {Gezari}, Suvi and {Miller-Jones}, James C.~A. and {Frederick}, Sara and {Sfaradi}, Itai and {Bietenholz}, Michael F. and {Horesh}, Assaf and {Fender}, Rob and {Garrappa}, Simone and {Ahumada}, Tom{\'a}s and {Andreoni}, Igor and {Belicki}, Justin and {Bellm}, Eric C. and {B{\"o}ttcher}, Markus and {Brinnel}, Valery and {Burruss}, Rick and {Cenko}, S. Bradley and {Coughlin}, Michael W. and {Cunningham}, Virginia and {Drake}, Andrew and {Farrar}, Glennys R. and {Feeney}, Michael and {Foley}, Ryan J. and {Gal-Yam}, Avishay and {Golkhou}, V. Zach and {Goobar}, Ariel and {Graham}, Matthew J. and {Hammerstein}, Erica and {Helou}, George and {Hung}, Tiara and {Kasliwal}, Mansi M. and {Kilpatrick}, Charles D. and {Kong}, Albert K.~H. and {Kupfer}, Thomas and {Laher}, Russ R. and {Mahabal}, Ashish A. and {Masci}, Frank J. and {Necker}, Jannis and {Nordin}, Jakob and {Perley}, Daniel A. and {Rigault}, Mickael and {Reusch}, Simeon and {Rodriguez}, Hector and {Rojas-Bravo}, C{\'e}sar and {Rusholme}, Ben and {Shupe}, David L. and {Singer}, Leo P. and {Sollerman}, Jesper and {Soumagnac}, Maayane T. and {Stern}, Daniel and {Taggart}, Kirsty and {van Santen}, Jakob and {Ward}, Charlotte and {Woudt}, Patrick and {Yao}, Yuhan},
        title = "{A tidal disruption event coincident with a high-energy neutrino}",
      journal = {Nature Astronomy},
         year = 2021,
        month = feb,
       volume = {5},
        pages = {510-518},
          doi = {10.1038/s41550-020-01295-8},
archivePrefix = {arXiv},
       eprint = {2005.05340},
 primaryClass = {astro-ph.HE},
       adsurl = {https://ui.adsabs.harvard.edu/abs/2021NatAs...5..510S}
}

@ARTICLE{2025NatAs...9..564L,
       author = {{Liu}, Y. and {Sun}, H. and {Xu}, D. and {Svinkin}, D.~S. and {Delaunay}, J. and {Tanvir}, N.~R. and {Gao}, H. and {Zhang}, C. and {Chen}, Y. and {Wu}, X.-F. and {Zhang}, B. and {Yuan}, W. and {An}, J. and {Bruni}, G. and {Frederiks}, D.~D. and {Ghirlanda}, G. and {Hu}, J.-W. and {Li}, A. and {Li}, C.-K. and {Li}, J.-D. and {Malesani}, D.~B. and {Piro}, L. and {Raman}, G. and {Ricci}, R. and {Troja}, E. and {Vergani}, S.~D. and {Wu}, Q.-Y. and {Yang}, J. and {Zhang}, B.-B. and {Zhu}, Z.-P. and {de Ugarte Postigo}, A. and {Demin}, A.~G. and {Dobie}, D. and {Fan}, Z. and {Fu}, S.-Y. and {Fynbo}, J.~P.~U. and {Geng}, J.-J. and {Gianfagna}, G. and {Hu}, Y.-D. and {Huang}, Y.-F. and {Jiang}, S.-Q. and {Jonker}, P.~G. and {Julakanti}, Y. and {Kennea}, J.~A. and {Kokomov}, A.~A. and {Kuulkers}, E. and {Lei}, W.-H. and {Leung}, J.~K. and {Levan}, A.~J. and {Li}, D.-Y. and {Li}, Y. and {Littlefair}, S.~P. and {Liu}, X. and {Lysenko}, A.~L. and {Ma}, Y.-N. and {Martin-Carrillo}, A. and {O'Brien}, P. and {Parsotan}, T. and {Quirola-V{\'a}squez}, J. and {Ridnaia}, A.~V. and {Ronchini}, S. and {Rossi}, A. and {Mata-S{\'a}nchez}, D. and {Schneider}, B. and {Shen}, R.-F. and {Thakur}, A.~L. and {Tohuvavohu}, A. and {Torres}, M.~A.~P. and {Tsvetkova}, A.~E. and {Ulanov}, M.~V. and {Wei}, J.-J. and {Xiao}, D. and {Yin}, Y.-H.~I. and {Bai}, M. and {Burwitz}, V. and {Cai}, Z.-M. and {Chen}, F.-S. and {Chen}, H.-L. and {Chen}, T.-X. and {Chen}, W. and {Chen}, Y.-F. and {Chen}, Y.-H. and {Cheng}, H.-Q. and {Cordier}, B. and {Cui}, C.-Z. and {Cui}, W.-W. and {Dai}, Y.-F. and {Dai}, Z.-G. and {Eder}, J. and {Eyles-Ferris}, R.~A.~J. and {Fan}, D.-W. and {Feldman}, C. and {Feng}, H. and {Feng}, Z. and {Friedrich}, P. and {Gao}, X. and {Gonzalez}, J.-F. and {Guan}, J. and {Han}, D.-W. and {Han}, J. and {Hou}, D.-J. and {Hu}, H.-B. and {Hu}, T. and {Huang}, M.-H. and {Huo}, J. and {Hutchinson}, I. and {Ji}, Z. and {Jia}, S.-M. and {Jia}, Z.-Q. and {Jiang}, B.-W. and {Jin}, C.-C. and {Jin}, G. and {Jin}, J.-J. and {Keereman}, A. and {Lerman}, H. and {Li}, J.-F. and {Li}, L.-H. and {Li}, M.-S. and {Li}, W. and {Li}, Z.-D. and {Lian}, T.-Y. and {Liang}, E.-W. and {Ling}, Z.-X. and {Liu}, C.-Z. and {Liu}, H.-Y. and {Liu}, H.-Q. and {Liu}, M.-J. and {Liu}, Y.-R. and {Lu}, F.-J. and {L{\"u}}, H.-J. and {Luo}, L.-D. and {Ma}, F.~L. and {Ma}, J. and {Mao}, J.-R. and {Mao}, X. and {McHugh}, M. and {Meidinger}, N. and {Nandra}, K. and {Osborne}, J.~P. and {Pan}, H.-W. and {Pan}, X. and {Ravasio}, M.~E. and {Rau}, A. and {Rea}, N. and {Rehman}, U. and {Sanders}, J. and {Santovincenzo}, A. and {Song}, L.-M. and {Su}, J. and {Sun}, L.-J. and {Sun}, S.-L. and {Sun}, X.-J. and {Tan}, Y.-Y. and {Tang}, Q.-J. and {Tao}, Y.-H. and {Tong}, J.-Z. and {Wang}, C.-Y. and {Wang}, H. and {Wang}, J. and {Wang}, L. and {Wang}, W.-X. and {Wang}, X.-F. and {Wang}, X.-Y. and {Wang}, Y.-L. and {Wang}, Y.-S. and {Wei}, D.-M. and {Willingale}, R. and {Xiong}, S.-L. and {Xu}, H.-T. and {Xu}, J.-J. and {Xu}, X.-P. and {Xu}, Y.-F. and {Xu}, Z. and {Xue}, C.-B. and {Xue}, Y.-L. and {Yan}, A.-L. and {Yang}, F. and {Yang}, H.-N. and {Yang}, X.-T. and {Yang}, Y.-J. and {Yu}, Y.-W. and {Zhang}, J. and {Zhang}, M. and {Zhang}, S.-N. and {Zhang}, W.-D. and {Zhang}, W.-J. and {Zhang}, Y.-H. and {Zhang}, Z. and {Zhang}, Z. and {Zhang}, Z.-L. and {Zhao}, D.-H. and {Zhao}, H.-S. and {Zhao}, X.-F. and {Zhao}, Z.-J. and {Zhou}, L.-X. and {Zhou}, Y.-L. and {Zhu}, Y.-X. and {Zhu}, Z.-C. and {Zuo}, X.-X.},
        title = "{Soft X-ray prompt emission from the high-redshift gamma-ray burst EP240315a}",
      journal = {Nature Astronomy},
         year = 2025,
        month = apr,
       volume = {9},
        pages = {564-576},
          doi = {10.1038/s41550-024-02449-8},
archivePrefix = {arXiv},
       eprint = {2404.16425},
 primaryClass = {astro-ph.HE},
       adsurl = {https://ui.adsabs.harvard.edu/abs/2025NatAs...9..564L}
}

@ARTICLE{2026MNRAS.546ag046S,
       author = {{Stephens}, I. and {Rhodes}, L. and {Cooper}, A.~J. and {Motta}, S.~E. and {Bright}, J.~S.},
        title = "{Exploring the potential for ultra-relativistic jets in Scorpius X-1 with low angular resolution radio observations}",
      journal = {\mnras},
         year = 2026,
        month = feb,
       volume = {546},
       number = {2},
          eid = {stag046},
        pages = {stag046},
          doi = {10.1093/mnras/stag046},
archivePrefix = {arXiv},
       eprint = {2601.03962},
 primaryClass = {astro-ph.HE},
       adsurl = {https://ui.adsabs.harvard.edu/abs/2026MNRAS.546ag046S}
}

@ARTICLE{fender_1999,
       author = {{Fender}, R.~P. and {Garrington}, S.~T. and {McKay}, D.~J. and {Muxlow}, T.~W.~B. and {Pooley}, G.~G. and {Spencer}, R.~E. and {Stirling}, A.~M. and {Waltman}, E.~B.},
        title = "{MERLIN observations of relativistic ejections from GRS 1915+105}",
      journal = {\mnras},
         year = 1999,
        month = apr,
       volume = {304},
       number = {4},
        pages = {865-876},
          doi = {10.1046/j.1365-8711.1999.02364.x},
archivePrefix = {arXiv},
       eprint = {astro-ph/9812150},
 primaryClass = {astro-ph},
       adsurl = {https://ui.adsabs.harvard.edu/abs/1999MNRAS.304..865F}
}

@ARTICLE{Hjellming_1981,
       author = {{Hjellming}, R.~M. and {Johnston}, K.~J.},
        title = "{An analysis of the proper motions of SS 433 radio jets.}",
      journal = {\apjl},
         year = 1981,
        month = jun,
       volume = {246},
        pages = {L141-L145},
          doi = {10.1086/183571},
       adsurl = {https://ui.adsabs.harvard.edu/abs/1981ApJ...246L.141H}
}

@ARTICLE{cowie_2025,
       author = {{Cowie}, F.~J. and {Fender}, R.~P. and {Heywood}, I. and {Hughes}, A.~K. and {Savard}, K. and {Woudt}, P.~A. and {Carotenuto}, F. and {Cooper}, A.~J. and {van den Eijnden}, J. and {Gasealahwe}, K.~V.~S. and {Motta}, S.~E. and {Saikia}, P.},
        title = "{Relativistic precessing jets powered by an accreting neutron star}",
      journal = {\mnras},
         year = 2025,
        month = nov,
       volume = {544},
       number = {1},
        pages = {L37-L44},
          doi = {10.1093/mnrasl/slaf097},
archivePrefix = {arXiv},
       eprint = {2509.08951},
 primaryClass = {astro-ph.HE},
       adsurl = {https://ui.adsabs.harvard.edu/abs/2025MNRAS.544L..37C}
}

@ARTICLE{2025ApJ...995...61S,
       author = {{Schroeder}, Genevieve and {Ho}, Anna Y.~Q. and {Dastidar}, Ranadeep G. and {Modjaz}, Maryam and {Corsi}, Alessandra and {Duffell}, Paul C.},
        title = "{A Late-time Radio Search for Highly Off-axis Jets from PTF Broad-lined Ic Supernovae in GRB-like Host Galaxy Environments}",
      journal = {\apj},
         year = 2025,
        month = dec,
       volume = {995},
       number = {1},
          eid = {61},
        pages = {61},
          doi = {10.3847/1538-4357/ae129b},
archivePrefix = {arXiv},
       eprint = {2507.15928},
 primaryClass = {astro-ph.HE},
       adsurl = {https://ui.adsabs.harvard.edu/abs/2025ApJ...995...61S}
}

@ARTICLE{2025ApJ...986..106G,
       author = {{Gao}, Hao-Xuan and {Geng}, Jin-Jun and {Liang}, Yi-Fang and {Sun}, Hui and {Xu}, Fan and {Wu}, Xue-Feng and {Huang}, Yong-Feng and {Dai}, Zi-Gao and {Yuan}, Wei-Min},
        title = "{The Soft X-Ray Aspect of Gamma-Ray Bursts in the Einstein Probe Era}",
      journal = {\apj},
         year = 2025,
        month = jun,
       volume = {986},
       number = {1},
          eid = {106},
        pages = {106},
          doi = {10.3847/1538-4357/adceb1},
archivePrefix = {arXiv},
       eprint = {2410.21687},
 primaryClass = {astro-ph.HE},
       adsurl = {https://ui.adsabs.harvard.edu/abs/2025ApJ...986..106G}
}

@ARTICLE{gezari_2021,
       author = {{Gezari}, Suvi},
        title = "{Tidal Disruption Events}",
      journal = {\araa},
         year = 2021,
        month = sep,
       volume = {59},
        pages = {21-58},
          doi = {10.1146/annurev-astro-111720-030029},
archivePrefix = {arXiv},
       eprint = {2104.14580},
 primaryClass = {astro-ph.HE},
       adsurl = {https://ui.adsabs.harvard.edu/abs/2021ARA&A..59...21G}
}

@ARTICLE{Hjellming_1995,
       author = {{Hjellming}, R.~M. and {Rupen}, M.~P.},
        title = "{Episodic ejection of relativistic jets by the X-ray transient GRO J1655 - 40}",
      journal = {\nat},
         year = 1995,
        month = jun,
       volume = {375},
       number = {6531},
        pages = {464-468},
          doi = {10.1038/375464a0},
       adsurl = {https://ui.adsabs.harvard.edu/abs/1995Natur.375..464H}
}

@ARTICLE{Lobanov_2005,
       author = {{Lobanov}, A.~P.},
        title = "{Resolution limits in astronomical images}",
      journal = {arXiv e-prints},
         year = 2005,
        month = mar,
          eid = {astro-ph/0503225},
        pages = {astro-ph/0503225},
          doi = {10.48550/arXiv.astro-ph/0503225},
archivePrefix = {arXiv},
       eprint = {astro-ph/0503225},
 primaryClass = {astro-ph},
       adsurl = {https://ui.adsabs.harvard.edu/abs/2005astro.ph..3225L}
}

@ARTICLE{lilje_2026,
       author = {{Lilje}, Clara and {Fender}, Rob and {Matthews}, James H.},
        title = "{Kinematics show consistency between stellar mass and supermassive black hole parent population jet speeds}",
      journal = {\mnras},
         year = 2026,
        month = feb,
       volume = {545},
       number = {4},
          eid = {staf2102},
        pages = {staf2102},
          doi = {10.1093/mnras/staf2102},
archivePrefix = {arXiv},
       eprint = {2511.19362},
 primaryClass = {astro-ph.HE},
       adsurl = {https://ui.adsabs.harvard.edu/abs/2026MNRAS.545f2102L}
}

@ARTICLE{bacon_2026,
       author = {{Bacon}, Nicolas J. and {Cooper}, Alex J. and {Kantzas}, Dimitrios and {Matthews}, James H. and {Fender}, Rob},
        title = "{Cosmic rays, {\ensuremath{\gamma}}-rays, and neutrinos from discrete black hole X-ray binary ejecta}",
      journal = {\mnras},
         year = 2026,
        month = feb,
       volume = {546},
       number = {2},
          eid = {stag080},
        pages = {stag080},
          doi = {10.1093/mnras/stag080},
archivePrefix = {arXiv},
       eprint = {2601.07589},
 primaryClass = {astro-ph.HE},
       adsurl = {https://ui.adsabs.harvard.edu/abs/2026MNRAS.546ag080B}
}

@ARTICLE{carotenuto_2021,
       author = {{Carotenuto}, F. and {Corbel}, S. and {Tremou}, E. and {Russell}, T.~D. and {Tzioumis}, A. and {Fender}, R.~P. and {Woudt}, P.~A. and {Motta}, S.~E. and {Miller-Jones}, J.~C.~A. and {Chauhan}, J. and {Tetarenko}, A.~J. and {Sivakoff}, G.~R. and {Heywood}, I. and {Horesh}, A. and {van der Horst}, A.~J. and {Koerding}, E. and {Mooley}, K.~P.},
        title = "{The black hole transient MAXI J1348-630: evolution of the compact and transient jets during its 2019/2020 outburst}",
      journal = {\mnras},
         year = 2021,
        month = jun,
       volume = {504},
       number = {1},
        pages = {444-468},
          doi = {10.1093/mnras/stab864},
archivePrefix = {arXiv},
       eprint = {2103.12190},
 primaryClass = {astro-ph.HE},
       adsurl = {https://ui.adsabs.harvard.edu/abs/2021MNRAS.504..444C}
}

@ARTICLE{Tao_2018,
       author = {{Tao}, Lian and {Chen}, YuPeng and {G{\"u}ng{\"o}r}, Can and {Huang}, Yue and {Lu}, FangJun and {Qu}, JinLu and {Song}, LiMing and {Zhang}, Liang and {Zhang}, Shu and {Zhang}, ShuangNan},
        title = "{Swift observations of the bright uncatalogued X-ray transient MAXI J1535-571}",
      journal = {\mnras},
         year = 2018,
        month = nov,
       volume = {480},
       number = {4},
        pages = {4443-4454},
          doi = {10.1093/mnras/sty2157},
archivePrefix = {arXiv},
       eprint = {1806.08024},
 primaryClass = {astro-ph.HE},
       adsurl = {https://ui.adsabs.harvard.edu/abs/2018MNRAS.480.4443T}
}

@ARTICLE{huang_2018,
       author = {{Huang}, Y. and {Qu}, J.~L. and {Zhang}, S.~N. and {Bu}, Q.~C. and {Chen}, Y.~P. and {Tao}, L. and {Zhang}, S. and {Lu}, F.~J. and {Li}, T.~P. and {Song}, L.~M. and {Xu}, Y.~P. and {Cao}, X.~L. and {Chen}, Y. and {Liu}, C.~Z. and {Chang}, H. -K. and {Yu}, W.~F. and {Weng}, S.~S. and {Hou}, X. and {Kong}, A.~K.~H. and {Xie}, F.~G. and {Zhang}, G.~B. and {ZHOU}, J.~F. and {Chang}, Z. and {Chen}, G. and {Chen}, L. and {Chen}, T.~X. and {Chen}, Y.~B. and {Cui}, W. and {Cui}, W.~W. and {Deng}, J.~K. and {Dong}, Y.~W. and {Du}, Y.~Y. and {Fu}, M.~X. and {Gao}, G.~H. and {Gao}, H. and {Gao}, M. and {Ge}, M.~Y. and {Gu}, Y.~D. and {Guan}, J. and {Gungor}, C. and {Guo}, C.~C. and {Han}, D.~W. and {Hu}, W. and {Huo}, J. and {Ji}, J.~F. and {Jia}, S.~M. and {Jiang}, L.~H. and {Jiang}, W.~C. and {Jin}, J. and {Jin}, Y.~J. and {Li}, B. and {Li}, C.~K. and {Li}, G. and {Li}, M.~S. and {Li}, W. and {Li}, X. and {Li}, X.~B. and {Li}, X.~F. and {Li}, Y.~G. and {Li}, Z.~J. and {Li}, Z.~W. and {Liang}, X.~H. and {Liao}, J.~Y. and {Liu}, G.~Q. and {Liu}, H.~W. and {Liu}, S.~Z. and {Liu}, X.~J. and {Liu}, Y. and {Liu}, Y.~N. and {Lu}, B. and {Lu}, X.~F. and {Luo}, T. and {Ma}, X. and {Meng}, B. and {Nang}, Y. and {Nie}, J.~Y. and {Ou}, G. and {Sai}, N. and {Shang}, R.~C. and {Sun}, L. and {Tan}, Y. and {Tao}, W. and {Tuo}, Y.~L. and {Wang}, G.~F. and {Wang}, H.~Y. and {Wang}, J. and {Wang}, W.~S. and {Wang}, Y.~S. and {Wen}, X.~Y. and {Wu}, B.~B. and {Wu}, M. and {Xiao}, G.~C. and {Xiong}, S.~L. and {Xu}, H. and {Yan}, L.~L. and {Yang}, J.~W. and {Yang}, S. and {Yang}, Y.~J. and {Zhang}, A.~M. and {Zhang}, C.~L. and {Zhang}, C.~M. and {Zhang}, F. and {Zhang}, H.~M. and {Zhang}, J. and {Zhang}, Q. and {Zhang}, T. and {Zhang}, W. and {Zhang}, W.~C. and {Zhang}, W.~Z. and {Zhang}, Y. and {Zhang}, Y. and {Zhang}, Y.~F. and {Zhang}, Y.~J. and {Zhang}, Z. and {Zhang}, Z. and {Zhang}, Z.~L. and {Zhao}, H.~S. and {Zhao}, J.~L. and {Zhao}, X.~F. and {Zheng}, S.~J. and {Zhu}, Y. and {Zhu}, Y.~X. and {Zou}, C.~L. and {Insight-HXMT Collaboration}},
        title = "{INSIGHT-HXMT Observations of the New Black Hole Candidate MAXI J1535-571: Timing Analysis}",
      journal = {\apj},
         year = 2018,
        month = oct,
       volume = {866},
       number = {2},
          eid = {122},
        pages = {122},
          doi = {10.3847/1538-4357/aade4c},
archivePrefix = {arXiv},
       eprint = {1808.05318},
 primaryClass = {astro-ph.HE},
       adsurl = {https://ui.adsabs.harvard.edu/abs/2018ApJ...866..122H}
}

@ARTICLE{russell_2017,
       author = {{Russell}, T.~D. and {Miller-Jones}, J.~C.~A. and {Sivakoff}, G.~R. and {Tetarenko}, A.~J. and {Jacpot Xrb Collaboration}},
        title = "{ATCA radio detection of MAXI J1535-571 indicates it is a strong black hole X-ray binary candidate}",
      journal = {The Astronomer's Telegram},
         year = 2017,
        month = sep,
       volume = {10711},
        pages = {1},
       adsurl = {https://ui.adsabs.harvard.edu/abs/2017ATel10711....1R}
}

@ARTICLE{zhang_2025,
       author = {{Zhang}, Xian and {Yu}, Wenfei and {Carotenuto}, Francesco and {Fender}, Rob and {Motta}, Sara and {Bahramian}, Arash and {Miller-Jones}, James C.~A. and {Russell}, Thomas D. and {Corbel}, Stephane and {Woudt}, Patrick A. and {Atri}, Pikky and {Knigge}, Christian and {Sivakoff}, Gregory R. and {Hughes}, Andrew K. and {van den Eijnden}, Jakob and {Matthews}, James and {Baglio}, Maria C. and {Saikia}, Payaswini},
        title = "{Jets from a stellar-mass black hole are as relativistic as those from supermassive black holes}",
      journal = {arXiv e-prints},
         year = 2025,
        month = apr,
          eid = {arXiv:2504.11945},
        pages = {arXiv:2504.11945},
          doi = {10.48550/arXiv.2504.11945},
archivePrefix = {arXiv},
       eprint = {2504.11945},
 primaryClass = {astro-ph.HE},
       adsurl = {https://ui.adsabs.harvard.edu/abs/2025arXiv250411945Z}
}

@article{savard_2025,
       author = {{Savard}, Katie and {Matthews}, James H. and {Fender}, Rob and {Heywood}, Ian},
        title = "{Relativistic ejecta from stellar mass black holes: insights from simulations and synthetic radio images}",
      journal = {\mnras},
         year = 2025,
        month = jun,
       volume = {540},
       number = {1},
        pages = {1084-1106},
          doi = {10.1093/mnras/staf739},
archivePrefix = {arXiv},
       eprint = {2504.20914},
 primaryClass = {astro-ph.HE},
       adsurl = {https://ui.adsabs.harvard.edu/abs/2025MNRAS.540.1084S}
}

@ARTICLE{steiner_2012,
       author = {{Steiner}, James F. and {McClintock}, Jeffrey E.},
        title = "{Modeling the Jet Kinematics of the Black Hole Microquasar XTE J1550-564: A Constraint on Spin-Orbit Alignment}",
      journal = {\apj},
         year = 2012,
        month = feb,
       volume = {745},
       number = {2},
          eid = {136},
        pages = {136},
          doi = {10.1088/0004-637X/745/2/136},
archivePrefix = {arXiv},
       eprint = {1110.6849},
 primaryClass = {astro-ph.HE},
       adsurl = {https://ui.adsabs.harvard.edu/abs/2012ApJ...745..136S}
}

@ARTICLE{wood_2023,
       author = {{Wood}, C.~M. and {Miller-Jones}, J.~C.~A. and {Bahramian}, A. and {Tingay}, S.~J. and {Russell}, T.~D. and {Tetarenko}, A.~J. and {Altamirano}, D. and {Belloni}, T. and {Carotenuto}, F. and {Ceccobello}, C. and {Corbel}, S. and {Espinasse}, M. and {Fender}, R.~P. and {K{\"o}rding}, E. and {Migliari}, S. and {Russell}, D.~M. and {Sarazin}, C.~L. and {Sivakoff}, G.~R. and {Soria}, R. and {Tudose}, V.},
        title = "{Time-dependent visibility modelling of a relativistic jet in the X-ray binary MAXI J1803-298}",
      journal = {\mnras},
         year = 2023,
        month = jun,
       volume = {522},
       number = {1},
        pages = {70-89},
          doi = {10.1093/mnras/stad939},
archivePrefix = {arXiv},
       eprint = {2303.15648},
 primaryClass = {astro-ph.HE},
       adsurl = {https://ui.adsabs.harvard.edu/abs/2023MNRAS.522...70W}
}

@ARTICLE{bahramian_2023,
       author = {{Bahramian}, A. and {Tremou}, E. and {Tetarenko}, A.~J. and {Miller-Jones}, J.~C.~A. and {Fender}, R.~P. and {Corbel}, S. and {Williams}, D.~R.~A. and {Strader}, J. and {Carotenuto}, F. and {Salinas}, R. and {Kennea}, J.~A. and {Motta}, S.~E. and {Woudt}, P.~A. and {Matthews}, J.~H. and {Russell}, T.~D.},
        title = "{MAXI J1848-015: The First Detection of Relativistically Moving Outflows from a Globular Cluster X-Ray Binary}",
      journal = {\apjl},
         year = 2023,
        month = may,
       volume = {948},
       number = {1},
          eid = {L7},
        pages = {L7},
          doi = {10.3847/2041-8213/accde1},
archivePrefix = {arXiv},
       eprint = {2305.03764},
 primaryClass = {astro-ph.HE},
       adsurl = {https://ui.adsabs.harvard.edu/abs/2023ApJ...948L...7B}
}

@ARTICLE{williams_2022,
       author = {{Williams}, D.~R.~A. and {Motta}, S.~E. and {Fender}, R. and {Miller-Jones}, J.~C.~A. and {Neilsen}, J. and {Allison}, J.~R. and {Bright}, J. and {Heywood}, I. and {Jacob}, P.~F.~L. and {Rhodes}, L. and {Tremou}, E. and {Woudt}, P.~A. and {Eijnden}, J. van den and {Carotenuto}, F. and {Green}, D.~A. and {Titterington}, D. and {van der Horst}, A.~J. and {Saikia}, P.},
        title = "{Radio observations of the Black Hole X-ray Binary EXO 1846-031 re-awakening from a 34-year slumber}",
      journal = {\mnras},
         year = 2022,
        month = dec,
       volume = {517},
       number = {2},
        pages = {2801-2817},
          doi = {10.1093/mnras/stac2700},
archivePrefix = {arXiv},
       eprint = {2209.10228},
 primaryClass = {astro-ph.HE},
       adsurl = {https://ui.adsabs.harvard.edu/abs/2022MNRAS.517.2801W}
}

@ARTICLE{bright_2020,
       author = {{Bright}, J.~S. and {Fender}, R.~P. and {Motta}, S.~E. and {Williams}, D.~R.~A. and {Moldon}, J. and {Plotkin}, R.~M. and {Miller-Jones}, J.~C.~A. and {Heywood}, I. and {Tremou}, E. and {Beswick}, R. and {Sivakoff}, G.~R. and {Corbel}, S. and {Buckley}, D.~A.~H. and {Homan}, J. and {Gallo}, E. and {Tetarenko}, A.~J. and {Russell}, T.~D. and {Green}, D.~A. and {Titterington}, D. and {Woudt}, P.~A. and {Armstrong}, R.~P. and {Groot}, P.~J. and {Horesh}, A. and {van der Horst}, A.~J. and {K{\"o}rding}, E.~G. and {McBride}, V.~A. and {Rowlinson}, A. and {Wijers}, R.~A.~M.~J.},
        title = "{An extremely powerful long-lived superluminal ejection from the black hole MAXI J1820+070}",
      journal = {Nature Astronomy},
         year = 2020,
        month = mar,
       volume = {4},
        pages = {697-703},
          doi = {10.1038/s41550-020-1023-5},
archivePrefix = {arXiv},
       eprint = {2003.01083},
 primaryClass = {astro-ph.HE},
       adsurl = {https://ui.adsabs.harvard.edu/abs/2020NatAs...4..697B}
}

@ARTICLE{corbel_2002,
       author = {{Corbel}, S. and {Fender}, R.~P. and {Tzioumis}, A.~K. and {Tomsick}, J.~A. and {Orosz}, J.~A. and {Miller}, J.~M. and {Wijnands}, R. and {Kaaret}, P.},
        title = "{Large-Scale, Decelerating, Relativistic X-ray Jets from the Microquasar XTE J1550-564}",
      journal = {Science},
         year = 2002,
        month = oct,
       volume = {298},
       number = {5591},
        pages = {196-199},
          doi = {10.1126/science.1075857},
archivePrefix = {arXiv},
       eprint = {astro-ph/0210224},
 primaryClass = {astro-ph},
       adsurl = {https://ui.adsabs.harvard.edu/abs/2002Sci...298..196C}
}

@ARTICLE{russell_2019,
       author = {{Russell}, T.~D. and {Tetarenko}, A.~J. and {Miller-Jones}, J.~C.~A. and {Sivakoff}, G.~R. and {Parikh}, A.~S. and {Rapisarda}, S. and {Wijnands}, R. and {Corbel}, S. and {Tremou}, E. and {Altamirano}, D. and {Baglio}, M.~C. and {Ceccobello}, C. and {Degenaar}, N. and {van den Eijnden}, J. and {Fender}, R. and {Heywood}, I. and {Krimm}, H.~A. and {Lucchini}, M. and {Markoff}, S. and {Russell}, D.~M. and {Soria}, R. and {Woudt}, P.~A.},
        title = "{Disk-Jet Coupling in the 2017/2018 Outburst of the Galactic Black Hole Candidate X-Ray Binary MAXI J1535-571}",
      journal = {\apj},
         year = 2019,
        month = oct,
       volume = {883},
       number = {2},
          eid = {198},
        pages = {198},
          doi = {10.3847/1538-4357/ab3d36},
archivePrefix = {arXiv},
       eprint = {1906.00998},
 primaryClass = {astro-ph.HE},
       adsurl = {https://ui.adsabs.harvard.edu/abs/2019ApJ...883..198R}
}

@ARTICLE{wood_2021,
       author = {{Wood}, C.~M. and {Miller-Jones}, J.~C.~A. and {Homan}, J. and {Bright}, J.~S. and {Motta}, S.~E. and {Fender}, R.~P. and {Markoff}, S. and {Belloni}, T.~M. and {K{\"o}rding}, E.~G. and {Maitra}, D. and {Migliari}, S. and {Russell}, D.~M. and {Russell}, T.~D. and {Sarazin}, C.~L. and {Soria}, R. and {Tetarenko}, A.~J. and {Tudose}, V.},
        title = "{The varying kinematics of multiple ejecta from the black hole X-ray binary MAXI J1820 + 070}",
      journal = {\mnras},
         year = 2021,
        month = aug,
       volume = {505},
       number = {3},
        pages = {3393-3403},
          doi = {10.1093/mnras/stab1479},
archivePrefix = {arXiv},
       eprint = {2105.09529},
 primaryClass = {astro-ph.HE},
       adsurl = {https://ui.adsabs.harvard.edu/abs/2021MNRAS.505.3393W}
}

@ARTICLE{mirabel_1994,
       author = {{Mirabel}, I.~F. and {Rodr{\'\i}guez}, L.~F.},
        title = "{A superluminal source in the Galaxy}",
      journal = {\nat},
         year = 1994,
        month = sep,
       volume = {371},
       number = {6492},
        pages = {46-48},
          doi = {10.1038/371046a0},
       adsurl = {https://ui.adsabs.harvard.edu/abs/1994Natur.371...46M}
}

@ARTICLE{mioduszewski_2001,
       author = {{Mioduszewski}, Amy J. and {Rupen}, Michael P. and {Hjellming}, Robert M. and {Pooley}, Guy G. and {Waltman}, Elizabeth B.},
        title = "{A One-sided Highly Relativistic Jet from Cygnus X-3}",
      journal = {\apj},
         year = 2001,
        month = jun,
       volume = {553},
       number = {2},
        pages = {766-775},
          doi = {10.1086/320965},
archivePrefix = {arXiv},
       eprint = {astro-ph/0102018},
 primaryClass = {astro-ph},
       adsurl = {https://ui.adsabs.harvard.edu/abs/2001ApJ...553..766M}
}

@ARTICLE{hannikainen_2001,
       author = {{Hannikainen}, Diana and {Campbell-Wilson}, Duncan and {Hunstead}, Richard and {McIntyre}, Vince and {Lovell}, Jim and {Reynolds}, John and {Tzioumis}, Tasso and {Wu}, Kinwah},
        title = "{XTE J1550{\textendash}564: a superluminal ejection during the September 1998 outburst}",
      journal = {Astrophysics and Space Science Supplement},
         year = 2001,
        month = jan,
       volume = {276},
        pages = {45-48},
          doi = {10.1023/A:1011659517584},
archivePrefix = {arXiv},
       eprint = {astro-ph/0011051},
 primaryClass = {astro-ph},
       adsurl = {https://ui.adsabs.harvard.edu/abs/2001ApSSS.276...45H}
}

@ARTICLE{corbel_2005,
       author = {{Corbel}, S. and {Kaaret}, P. and {Fender}, R.~P. and {Tzioumis}, A.~K. and {Tomsick}, J.~A. and {Orosz}, J.~A.},
        title = "{Discovery of X-Ray Jets in the Microquasar H1743-322}",
      journal = {\apj},
         year = 2005,
        month = oct,
       volume = {632},
       number = {1},
        pages = {504-513},
          doi = {10.1086/432499},
archivePrefix = {arXiv},
       eprint = {astro-ph/0505526},
 primaryClass = {astro-ph},
       adsurl = {https://ui.adsabs.harvard.edu/abs/2005ApJ...632..504C}
}

@ARTICLE{rushton_2017,
       author = {{Rushton}, A.~P. and {Miller-Jones}, J.~C.~A. and {Curran}, P.~A. and {Sivakoff}, G.~R. and {Rupen}, M.~P. and {Paragi}, Z. and {Spencer}, R.~E. and {Yang}, J. and {Altamirano}, D. and {Belloni}, T. and {Fender}, R.~P. and {Krimm}, H.~A. and {Maitra}, D. and {Migliari}, S. and {Russell}, D.~M. and {Russell}, T.~D. and {Soria}, R. and {Tudose}, V.},
        title = "{Resolved, expanding jets in the Galactic black hole candidate XTE J1908+094}",
      journal = {\mnras},
         year = 2017,
        month = jul,
       volume = {468},
       number = {3},
        pages = {2788-2802},
          doi = {10.1093/mnras/stx526},
archivePrefix = {arXiv},
       eprint = {1703.02110},
 primaryClass = {astro-ph.HE},
       adsurl = {https://ui.adsabs.harvard.edu/abs/2017MNRAS.468.2788R}
}

@ARTICLE{miller-jones_2019,
       author = {{Miller-Jones}, James C.~A. and {Tetarenko}, Alexandra J. and {Sivakoff}, Gregory R. and {Middleton}, Matthew J. and {Altamirano}, Diego and {Anderson}, Gemma E. and {Belloni}, Tomaso M. and {Fender}, Rob P. and {Jonker}, Peter G. and {K{\"o}rding}, Elmar G. and {Krimm}, Hans A. and {Maitra}, Dipankar and {Markoff}, Sera and {Migliari}, Simone and {Mooley}, Kunal P. and {Rupen}, Michael P. and {Russell}, David M. and {Russell}, Thomas D. and {Sarazin}, Craig L. and {Soria}, Roberto and {Tudose}, Valeriu},
        title = "{A rapidly changing jet orientation in the stellar-mass black-hole system V404 Cygni}",
      journal = {\nat},
         year = 2019,
        month = apr,
       volume = {569},
       number = {7756},
        pages = {374-377},
          doi = {10.1038/s41586-019-1152-0},
archivePrefix = {arXiv},
       eprint = {1906.05400},
 primaryClass = {astro-ph.HE},
       adsurl = {https://ui.adsabs.harvard.edu/abs/2019Natur.569..374M}
}

@ARTICLE{yang_2010,
       author = {{Yang}, J. and {Brocksopp}, C. and {Corbel}, S. and {Paragi}, Z. and {Tzioumis}, T. and {Fender}, R.~P.},
        title = "{A decelerating jet observed by the EVN and VLBA in the X-ray transient XTE J1752-223}",
      journal = {\mnras},
         year = 2010,
        month = nov,
       volume = {409},
       number = {1},
        pages = {L64-L68},
          doi = {10.1111/j.1745-3933.2010.00948.x},
archivePrefix = {arXiv},
       eprint = {1009.1367},
 primaryClass = {astro-ph.HE},
       adsurl = {https://ui.adsabs.harvard.edu/abs/2010MNRAS.409L..64Y}
}

@ARTICLE{gallo_2004,
       author = {{Gallo}, E. and {Corbel}, S. and {Fender}, R.~P. and {Maccarone}, T.~J. and {Tzioumis}, A.~K.},
        title = "{A transient large-scale relativistic radio jet from GX 339-4}",
      journal = {\mnras},
         year = 2004,
        month = jan,
       volume = {347},
       number = {3},
        pages = {L52-L56},
          doi = {10.1111/j.1365-2966.2004.07435.x},
archivePrefix = {arXiv},
       eprint = {astro-ph/0311452},
 primaryClass = {astro-ph},
       adsurl = {https://ui.adsabs.harvard.edu/abs/2004MNRAS.347L..52G}
}

@ARTICLE{hjellming_2000,
       author = {{Hjellming}, R.~M. and {Rupen}, M.~P. and {Hunstead}, R.~W. and {Campbell-Wilson}, D. and {Mioduszewski}, A.~J. and {Gaensler}, B.~M. and {Smith}, D.~A. and {Sault}, R.~J. and {Fender}, R.~P. and {Spencer}, R.~E. and {de la Force}, C.~J. and {Richards}, A.~M.~S. and {Garrington}, S.~T. and {Trushkin}, S.~A. and {Ghigo}, F.~D. and {Waltman}, E.~B. and {McCollough}, M.},
        title = "{Light Curves and Radio Structure of the 1999 September Transient Event in V4641 Sagittarii (=XTE J1819-254=SAX J1819.3-2525)}",
      journal = {\apj},
         year = 2000,
        month = dec,
       volume = {544},
       number = {2},
        pages = {977-992},
          doi = {10.1086/317255},
       adsurl = {https://ui.adsabs.harvard.edu/abs/2000ApJ...544..977H}
}

@ARTICLE{fender_2004,
       author = {{Fender}, R.~P. and {Belloni}, T.~M. and {Gallo}, E.},
        title = "{Towards a unified model for black hole X-ray binary jets}",
      journal = {\mnras},
         year = 2004,
        month = dec,
       volume = {355},
       number = {4},
        pages = {1105-1118},
          doi = {10.1111/j.1365-2966.2004.08384.x},
archivePrefix = {arXiv},
       eprint = {astro-ph/0409360},
 primaryClass = {astro-ph},
       adsurl = {https://ui.adsabs.harvard.edu/abs/2004MNRAS.355.1105F}
}

@ARTICLE{blandfordz_77,
       author = {{Blandford}, R.~D. and {Znajek}, R.~L.},
        title = "{Electromagnetic extraction of energy from Kerr black holes.}",
      journal = {\mnras},
         year = 1977,
        month = may,
       volume = {179},
        pages = {433-456},
          doi = {10.1093/mnras/179.3.433},
       adsurl = {https://ui.adsabs.harvard.edu/abs/1977MNRAS.179..433B}
}

@ARTICLE{nakahira_2018,
       author = {{Nakahira}, Satoshi and {Shidatsu}, Megumi and {Makishima}, Kazuo and {Ueda}, Yoshihiro and {Yamaoka}, Kazutaka and {Mihara}, Tatehiro and {Negoro}, Hitoshi and {Kawase}, Tomofumi and {Kawai}, Nobuyuki and {Morita}, Kotaro},
        title = "{Discovery and state transitions of the new Galactic black hole candidate MAXI J1535-571}",
      journal = {\pasj},
         year = 2018,
        month = oct,
       volume = {70},
       number = {5},
          eid = {95},
        pages = {95},
          doi = {10.1093/pasj/psy093},
archivePrefix = {arXiv},
       eprint = {1804.00800},
 primaryClass = {astro-ph.HE},
       adsurl = {https://ui.adsabs.harvard.edu/abs/2018PASJ...70...95N}
}

@ARTICLE{negoro_2017,
       author = {{Negoro}, H. and {Ishikawa}, M. and {Ueno}, S. and {Tomida}, H. and {Sugawara}, Y. and {Isobe}, N. and {Shimomukai}, R. and {Mihara}, T. and {Sugizaki}, M. and {Serino}, M. and {Iwakiri}, S. Nakahira W. and {Shidatsu}, M. and {Matsuoka}, M. and {Kawai}, N. and {Sugita}, S. and {Yoshii}, T. and {Tachibana}, Y. and {Harita}, S. and {Muraki}, Y. and {Morita}, K. and {Yoshida}, A. and {Sakamoto}, T. and {Kawakubo}, Y. and {Kitaoka}, Y. and {Hashimoto}, T. and {Tsunemi}, H. and {Yoneyama}, T. and {Nakajima}, M. and {Kawase}, T. and {Sakamaki}, A. and {Ueda}, Y. and {Hori}, T. and {Tanimoto}, A. and {Oda}, S. and {Tsuboi}, Y. and {Nakamura}, Y. and {Sasaki}, R. and {Kawai}, H. and {Yamauchi}, M. and {Hanyu}, C. and {Hidaka}, K. and {Kawamuro}, T. and {Yamaoka}, K.},
        title = "{MAXI/GSC discovery of a new hard X-ray transient MAXI J1535-571}",
      journal = {The Astronomer's Telegram},
         year = 2017,
        month = sep,
       volume = {10699},
        pages = {1},
       adsurl = {https://ui.adsabs.harvard.edu/abs/2017ATel10699....1N}
}

@ARTICLE{speagle_2020,
       author = {{Speagle}, Joshua S.},
        title = "{DYNESTY: a dynamic nested sampling package for estimating Bayesian posteriors and evidences}",
      journal = {\mnras},
         year = 2020,
        month = apr,
       volume = {493},
       number = {3},
        pages = {3132-3158},
          doi = {10.1093/mnras/staa278},
archivePrefix = {arXiv},
       eprint = {1904.02180},
 primaryClass = {astro-ph.IM},
       adsurl = {https://ui.adsabs.harvard.edu/abs/2020MNRAS.493.3132S}
}

@INPROCEEDINGS{Skilling_2004,
       author = {{Skilling}, John},
        title = "{Nested Sampling}",
    booktitle = {Bayesian Inference and Maximum Entropy Methods in Science and Engineering: 24th International Workshop on Bayesian Inference and Maximum Entropy Methods in Science and Engineering},
         year = 2004,
       editor = {{Fischer}, Rainer and {Preuss}, Roland and {Toussaint}, Udo Von},
       series = {American Institute of Physics Conference Series},
       volume = {735},
        month = nov,
    publisher = {AIP},
        pages = {395-405},
          doi = {10.1063/1.1835238},
       adsurl = {https://ui.adsabs.harvard.edu/abs/2004AIPC..735..395S}
}

@ARTICLE{espinasse_2020,
       author = {{Espinasse}, Mathilde and {Corbel}, St{\'e}phane and {Kaaret}, Philip and {Tremou}, Evangelia and {Migliori}, Giulia and {Plotkin}, Richard M. and {Bright}, Joe and {Tomsick}, John and {Tzioumis}, Anastasios and {Fender}, Rob and {Orosz}, Jerome A. and {Gallo}, Elena and {Homan}, Jeroen and {Jonker}, Peter G. and {Miller-Jones}, James C.~A. and {Russell}, David M. and {Motta}, Sara},
        title = "{Relativistic X-Ray Jets from the Black Hole X-Ray Binary MAXI J1820+070}",
      journal = {\apjl},
         year = 2020,
        month = jun,
       volume = {895},
       number = {2},
          eid = {L31},
        pages = {L31},
          doi = {10.3847/2041-8213/ab88b6},
archivePrefix = {arXiv},
       eprint = {2004.06416},
 primaryClass = {astro-ph.HE},
       adsurl = {https://ui.adsabs.harvard.edu/abs/2020ApJ...895L..31E}
}

@ARTICLE{foreman_mackey_2013,
       author = {{Foreman-Mackey}, Daniel and {Hogg}, David W. and {Lang}, Dustin and {Goodman}, Jonathan},
        title = "{emcee: The MCMC Hammer}",
      journal = {\pasp},
         year = 2013,
        month = mar,
       volume = {125},
       number = {925},
        pages = {306},
          doi = {10.1086/670067},
archivePrefix = {arXiv},
       eprint = {1202.3665},
 primaryClass = {astro-ph.IM},
       adsurl = {https://ui.adsabs.harvard.edu/abs/2013PASP..125..306F}
}

@ARTICLE{miller_2018,
       author = {{Miller}, J.~M. and {Gendreau}, K. and {Ludlam}, R.~M. and {Fabian}, A.~C. and {Altamirano}, D. and {Arzoumanian}, Z. and {Bult}, P.~M. and {Cackett}, E.~M. and {Homan}, J. and {Kara}, E. and {Neilsen}, J. and {Remillard}, R.~A. and {Tombesi}, F.},
        title = "{A NICER Spectrum of MAXI J1535-571: Near-maximal Black Hole Spin and Potential Disk Warping}",
      journal = {\apjl},
         year = 2018,
        month = jun,
       volume = {860},
       number = {2},
          eid = {L28},
        pages = {L28},
          doi = {10.3847/2041-8213/aacc61},
archivePrefix = {arXiv},
       eprint = {1806.04115},
 primaryClass = {astro-ph.HE},
       adsurl = {https://ui.adsabs.harvard.edu/abs/2018ApJ...860L..28M}
}

\appendix
\section{Additional Plots}

\begin{figure}
  \centering
{\includegraphics[width=.5\textwidth]{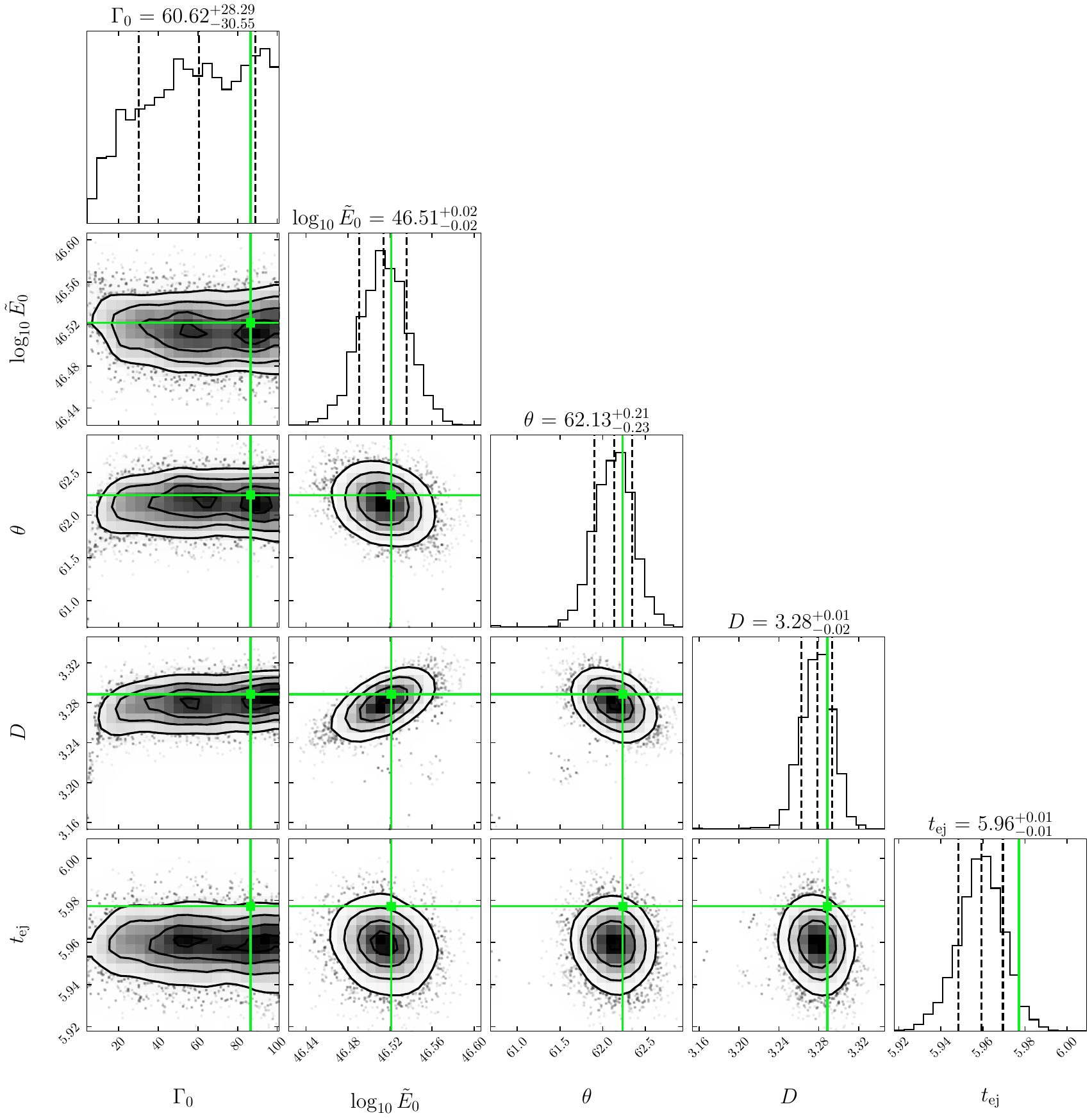}}
\caption{Corner plot of kinematic \texttt{MCMC} fit to MAXI J1820 referenced in Section \ref{sect:maxi1820}.}
\label{fig:1820_corner}
\end{figure}

\begin{figure}
  \centering
{\includegraphics[width=.5\textwidth]{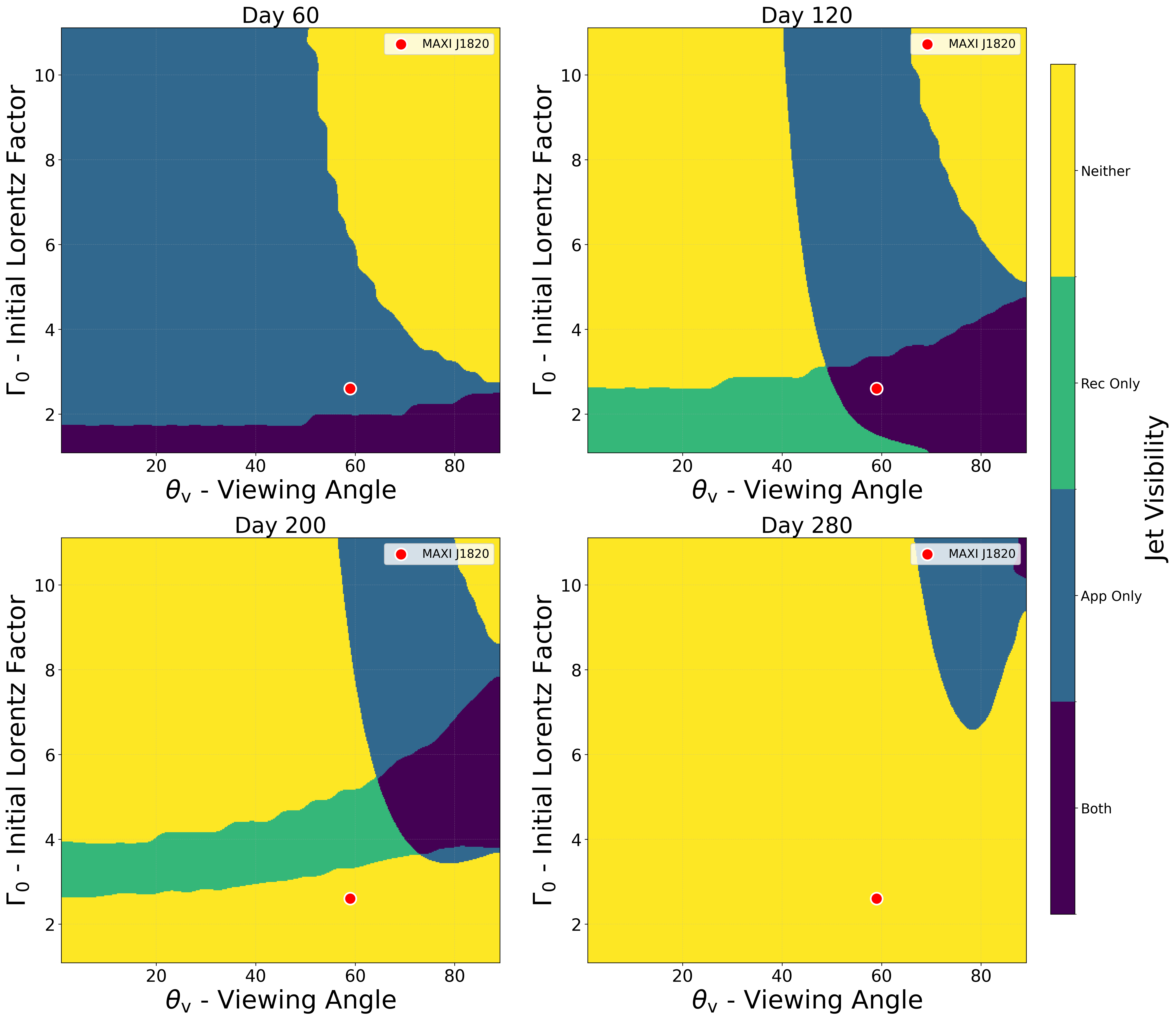}}
\caption{Detectability of MAXI J1820-like ejecta at 60, 120, 200, and 280 days post launch at $1.28$ GHz, as a function of the initial Lorentz factor $\Gamma_0$ and the viewing angle $\theta_{\rm v}$. Dark blue corresponds to regions of the parameter space where both jets are simultaneously detectable, lighter blue only the approaching jet, and green only the receding jet. The yellow region corresponds to the case when neither jet is above the $0.15 \, \rm mJy$ sensitivity threshold.}
\label{fig:observability}
\end{figure}





\bsp	
\label{lastpage}
\end{document}